\documentclass[useAMS,usenatbib]{mnras}
\usepackage[fleqn]{amsmath}

\usepackage{newtxtext,newtxmath}

\usepackage[T1]{fontenc}
\usepackage{ae,aecompl}

\usepackage{amsfonts,bm}
\usepackage{graphicx}
\usepackage[utf8]{inputenc}
\usepackage{adjustbox}   
\usepackage{subfig}
\usepackage{marvosym}

\newcommand{\tild}{~}

\newcommand  *{\diff}   {\mathop{}\!\mathrm{d}}
\renewcommand*{\vec}[1] {\boldsymbol{#1}}
\newcommand  *{\uvec}[1]{\hat{\vec{#1}}}
\newcommand  *{\s}[1]   {\mathsf{#1}}
\newcommand  *{\mat}[1] {\vec{\s{#1}}}
\newcommand  *{\Ups}    {\Upsilon}
\newcommand  *{\The}    {\Theta}
\newcommand  *{\I}      {\mathrm{i}}

\newcommand  *{\Exp}[1] {\mathrm{e}^{\textstyle #1}}
\newcommand  *{\phn}    {\phantom{-}}
\newcommand  *{\pfrac}[2] {\left(\frac{#1}{#2}\right)}
\newcommand   {\mbhm}{M_{\text{BH}}}
\newcommand   {\mbh}{$M_{\text{BH}}$}
\newcommand   {\ml}{$\Upsilon$}
\newcommand   {\mlm}{\Upsilon}

\title[Shape and anisotropy]{Accuracy and precision of triaxial orbit models II: Viewing angles, shape and orbital structure}

\author[S.~de Nicola et al.]{%
Stefano de Nicola,$^{\!\!1,2}$\thanks{E-mail: denicola@mpe.mpg.de}
Bianca Neureiter,$^{\!\!1,2}$
Jens Thomas,$^{\!\!1,2}$
Roberto P. Saglia,$^{\!\!1,2}$
\newauthor
and Ralf Bender $^{\!\!2,1}$
\\ \\
$^{1}$ Max-Planck Institute for Extraterrestrial Physics, Giessenbachstrasse 1, D-85748, Garching (Germany) \\
$^{2}$ Universit{\"a}ts-Sternwarte Muenchen, Scheinerstrasse 1, D-81679, Munich, Germany\\
}

\date{Accepted XXX. Received YYY; in original form ZZZ}

\pubyear{2021}

\begin{document}
\label{firstpage}
\pagerange{\pageref{firstpage}--\pageref{lastpage}}
\maketitle

\begin{abstract}

  We explore the potential of our novel triaxial modeling machinery in
recovering the viewing angles, the shape and the orbit distribution of
galaxies by using a high-resolution $N$-body merger simulation. Our
modelling technique includes several recent advancements. (i) Our new
triaxial deprojection algorithm SHAPE3D is able to significantly
shrink the range of possible orientations of a triaxial galaxy and
therefore to constrain its shape relying only on photometric
information. It also allows to probe degeneracies, i.e. to recover
different deprojections at the same assumed orientation.  With this
method we can constrain the intrinsic shape of the $N$-body
simulation, i.e. the axis ratios $p=b/a$ and $q=c/a$, with $\Delta p$
and $\Delta q$ $\lesssim$ 0.1 using only photometric information. The
typical accuracy of the viewing angles reconstruction is
15-20$^\circ$. (ii) Our new triaxial Schwarzschild code SMART exploits
the full kinematic information contained in the entire non-parametric
line-of-sight velocity distributions (LOSVDs) along with a 5D orbital
sampling in phase space. (iii) We use a new generalised information
criterion AIC$_p$ to optimise the smoothing and to select the best-fit
model, avoiding potential biases in purely $\chi^2$-based approaches.
With our deprojected densities, we recover the correct orbital
structure and anisotropy parameter $\beta$ with $\Delta \beta$
$\lesssim$ 0.1. These results are valid regardless of the tested
orientation of the simulation and suggest that even despite the known
intrinsic photometric and kinematic degeneracies the above described
advanced methods make it possible to recover the shape and the orbital
structure of triaxial bodies with unprecedented accuracy.

\end{abstract}




\section{Introduction}
\label{Sec.introduction}

The recovery of the intrinsic shape of a galaxy as well as the
reconstruction of the three-dimensional stellar dynamics rely on
projected quantities that we see on the plane of the sky. By using
2D-images of the galaxy on the plane of the sky, it is possible to
reconstruct the intrinsic 3D luminosity density (hereafter $\rho$)
that projects to the observed image (or isophotes) for a certain
galaxy inclination.
The shape of this 3D-density, measured in terms of the axis ratios $p \equiv y/x$, $q \equiv z/x$ and of the triaxiality parameter $T = (1 - p^2) / (1 - q^2)$ \citep{Franx91}, allows us to make inferences about the galaxy formation history. A particularly relevant example concerns the merging history that leads to the formation of elliptical galaxies: the most massive, triaxial, rounder galaxies, form through dry mergers, while fast-rotating, flatter galaxies, form through wet mergers (\citealt{Bender88b, Barnes92, Bender92, Kormendy96, Kormendy09b, Bois11,Khochfar11,Naab14}, see \citealt{Cappellari16} for a review). Moreover, the light density itself is used as a constraint to calculate the dynamics of
stars found around the galaxy center such that the resulting Line-of-Sight Velocity Distribution (LOSVD) matches the observed one
\citep{Sch79, Cretton99, Cretton00, Gebhardt00, Verolme02, Valluri04, Jens04, Valluri05, Jens05, VDB08, Vasiliev20, Bianca21}. With the stellar dynamics in hand, one can not only measure BH masses or mass-to-light (\ml) ratios, but also estimate the anisotropy profile $\beta (r)$. This is of particular relevance for core-elliptical galaxies, whose central cores are believed to be generated by BH-scouring and, hence, they should show a tangential bias in the innermost regions \citep{Jens14, Rantala18, Rantala19}. While dynamical models indeed suggested tangential anisotropy in the centers of elliptical galaxies of different kinds -- cored or non-cored (e.g. \citealt{Gebhardt03,Cappellari05,Schulze2011,McConnell2012,Jens14}) -- the structure of galaxies with depleted stellar cores is special: their anisotropy is remarkably uniform and changes from inner tangential anisotropy to outer radial anisotropy at the core radius as predicted (e.g. \citealt{Jens14, Jens16, Kianusch19}). \\
Generally, one uses both photometrical and kinematical data to measure the galaxy shapes. Nevertheless, regardless of the approach one chooses to tackle down the deprojection/dynamical modeling, knowing the galaxy inclination is a key ingredient. Unfortunately, the inclination cannot be measured in general, and the issue gets particularly severe when dealing with massive ellipticals, which are typically disk-less galaxies. Moreover, observational evidences such as isophotal twists or misalignment between kinematic and photometric axes
show that these galaxies are not axisymmetric, but rather triaxial \citep{Vincent2005, Ene18}, meaning that one needs to specify three viewing angles $\left(\theta, \phi, \psi\right)$ instead of only one angle $i$ needed in the
axisymmetric case \citep{Binney1985, deZeeuwFranx1989}. 
Assuming the wrong viewing angles when deprojecting a galaxy will
almost always yield the wrong shape \citep{dN20}, which will also
likely lead to a wrong estimate of the anisotropy profile $\beta(r)$ of the
galaxy and to biased estimates of the mass-to-light ratio \ml\,and the black hole mass \mbh\,when dynamically modeling the galaxy. \\
In principle, one could deproject a galaxy assuming a large
number of possible viewing angles and then dynamically model all
these three-dimensional luminosity densities to obtain an estimate of the galaxy viewing angles. In practice, this is not feasible because dynamical models are both very computing time- and memory-consuming. It is thus important to develop a deprojection tool which allows
for a significant reduction of the number of possible viewing angles,
generating physically plausible densities and keeping the degeneracy arising from the deprojection
under control \citep{Gerhard96, Kochanek96, Magorrian99, dN20}. One
commonly used routine, the Multi-Gaussian-Expansion (MGE,
\citealt{Cappellari02}) does not allow to explore different deprojections
for a given set of viewing angles and could in principle not yield a single
possible viewing angle if either the isophotes are very flattened
or the twist is large. On the dynamical modeling side, available triaxial
Schwarzschild codes (e.g. \citealt{VDB08}) deliver mass-to-light ratios and Dark Matter (DM) fractions of simulated galaxies deviating by 15-25\% from the true values (see \citealt{Jin19}).\\
Motivated by such arguments, we have developed two codes
aimed at filling these gaps:

\begin{itemize}
    \item In \citet{dN20} (hereafter dN20), we presented our
novel triaxial semi-parametric deprojection code SHAPE3D, that finds the best-
fit light density $\rho$ projecting to a certain surface brightness under the assumption of being strati-
fied onto “deformed” ellipsoids (see Sec. 3.1 of the paper). It also allows to bias the solution towards a certain degree of boxiness or diskiness and/or
to certain $p(r)$ and $q(r)$ profiles. Unless we observe a
triaxial galaxy exactly along one of the principal axes, the deprojection is unique if the density is a (deformed) ellipsoidal. Moreover, our code
tackles the degeneracy problem,
allowing for an exploration of possible density distributions that project to nothing. Finally, we
show how the possible viewing angles of a galaxy can
be significantly reduced from photometry alone. This not only helps the dynamical modeling, but also allows us to directly estimate galaxy shapes from photometry alone \citep{dN22}.
\item In \citet{Bianca21} (hereafter BN21), we presented our
novel triaxial Schwarzschild code SMART, which extends the axisymmetric
code of \citet{Jens04}. It exploits several advanced features: it fits the full non-parametric LOSVDs rather than Gauss-Hermite parametrisations, uses a 5-dimensional starting space for a better orbit sampling in the central regions and adopts a novel model selection technique to prevent overfitting, optimise the smoothing and deal with the different number of Degrees-of-Freedom (DOFs) for a given model \citep{Mathias21,Jens22}. The code was tested using a sophisticated, state-of-the-art $N$-body simulation aimed at reproducing the formation history of giant
ellipticals \citep{Rantala18, Rantala19} showing that if the intrinsic 3D distribution of the stars is known, the code shows excellent recoveries
of \mbh, \ml, the normalization of the DM halo as well as the internal
velocity moments with an accuracy never achieved by any pre-existing Schwarzschild codes.
\end{itemize}

Here we combine the two algorithms and use both photometry and
kinematics to constrain the galaxy viewing angles even better, and to
recover the correct galaxy shape and anisotropy profile. Moreover, we quantify
how large are the errors on the recovered mass parameters coming from the
deprojections. To our knowledge, only in \citet{VDB09} such study
has been attempted.  The paper
is structured as follows. In Sec. 2 we briefly report the main
features of the $N$-body simulation and our two codes. Sec. 3
describes our procedure, whose results are presented in Sec. 4.
Finally, we draw our conclusions in Sec. 5. Results on the recovery of
the mass parameter are presented in a companion paper (Neureiter
et al. 2022, submitted to MNRAS, hereafter Paper I).


\section{Data and Code}
\begin{figure*}

\subfloat[MIDDLE\label{Fig.Isophotes_middle}]{\includegraphics[width=.35\linewidth]{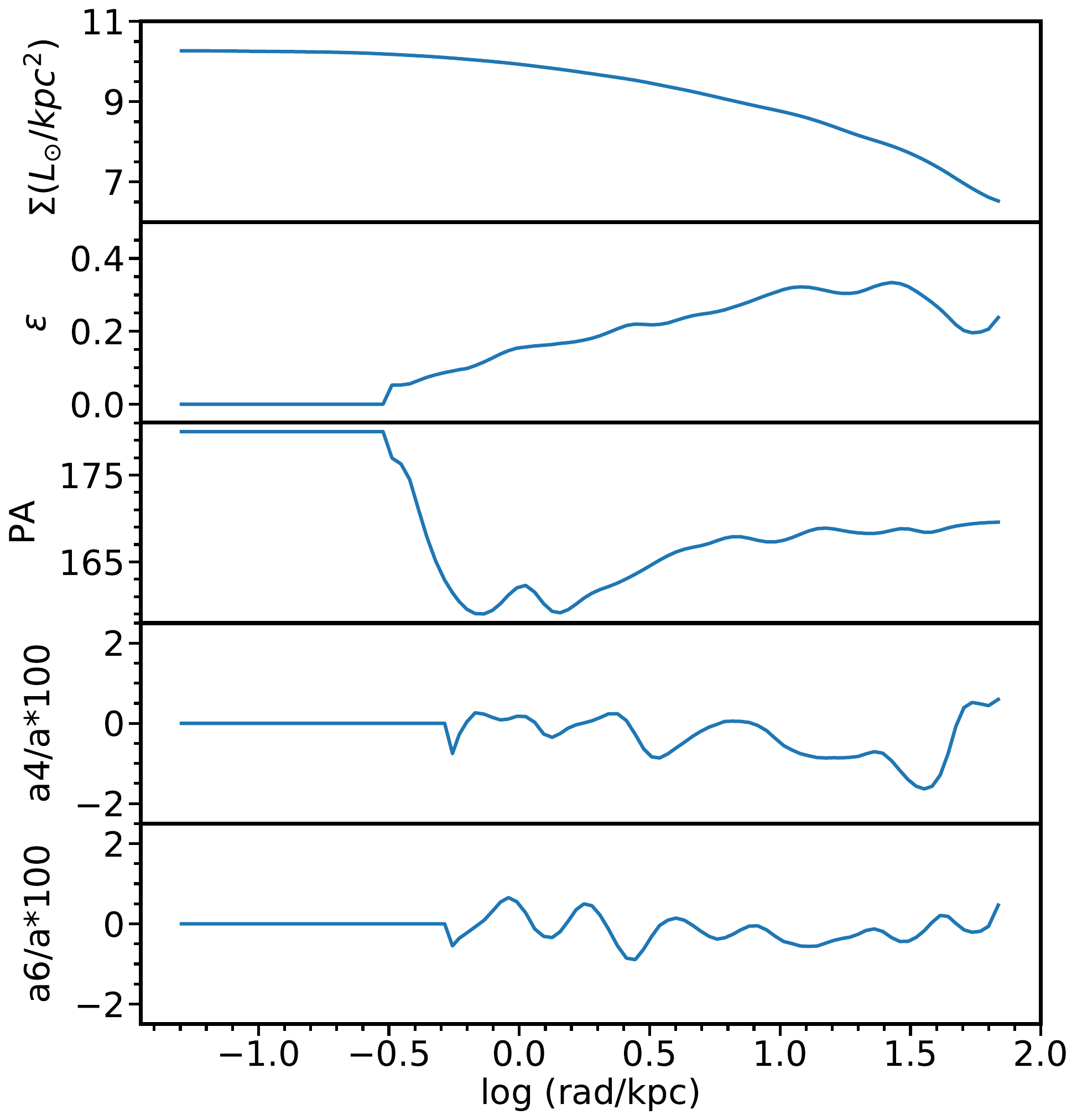}}
\subfloat[RAND\label{Fig.Isophotes_random}]{\includegraphics[width=.35\linewidth]{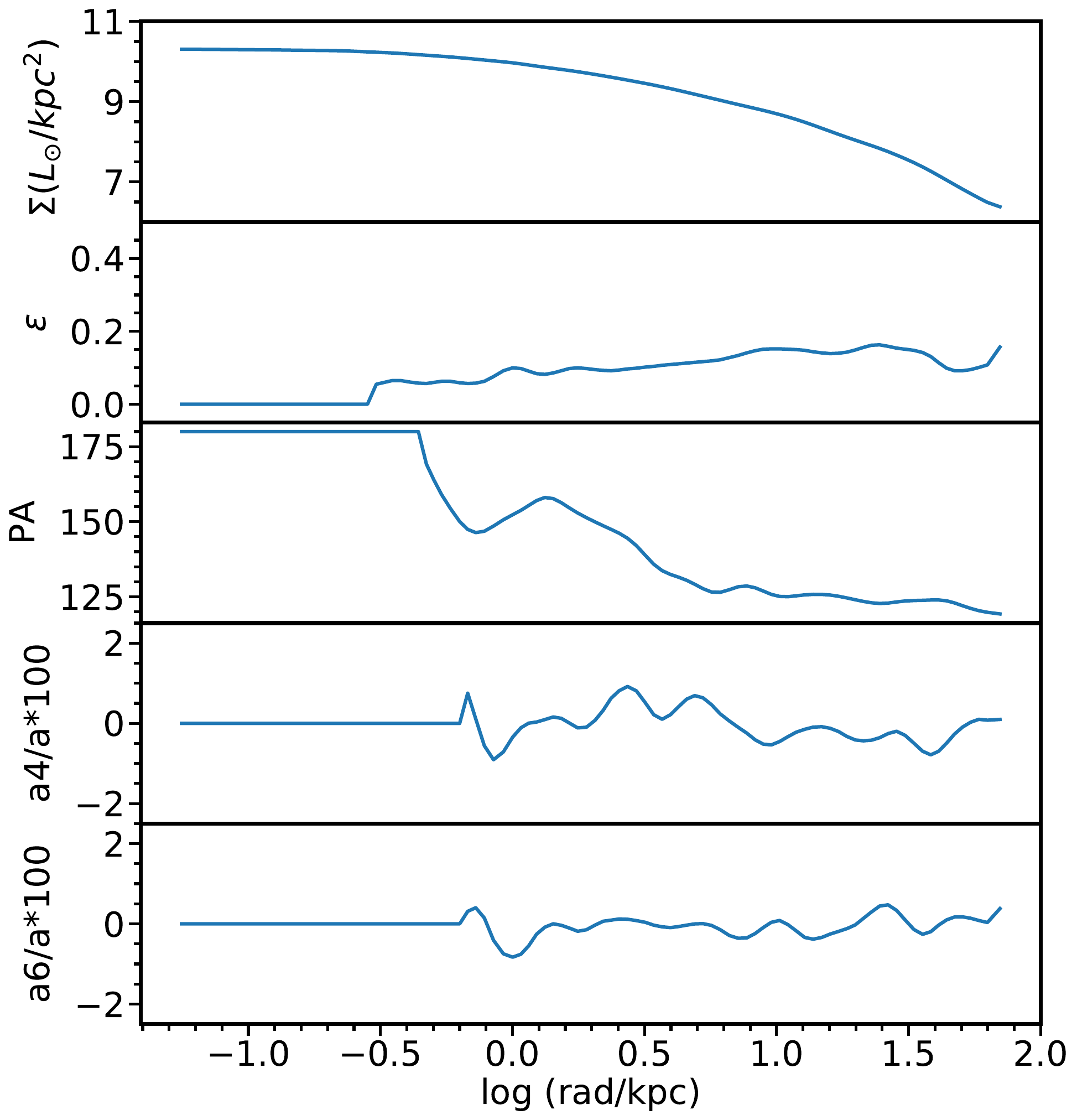}}

\subfloat[INTERM\label{Fig.Isophotes_intermediate}]{\includegraphics[width=.35\linewidth]{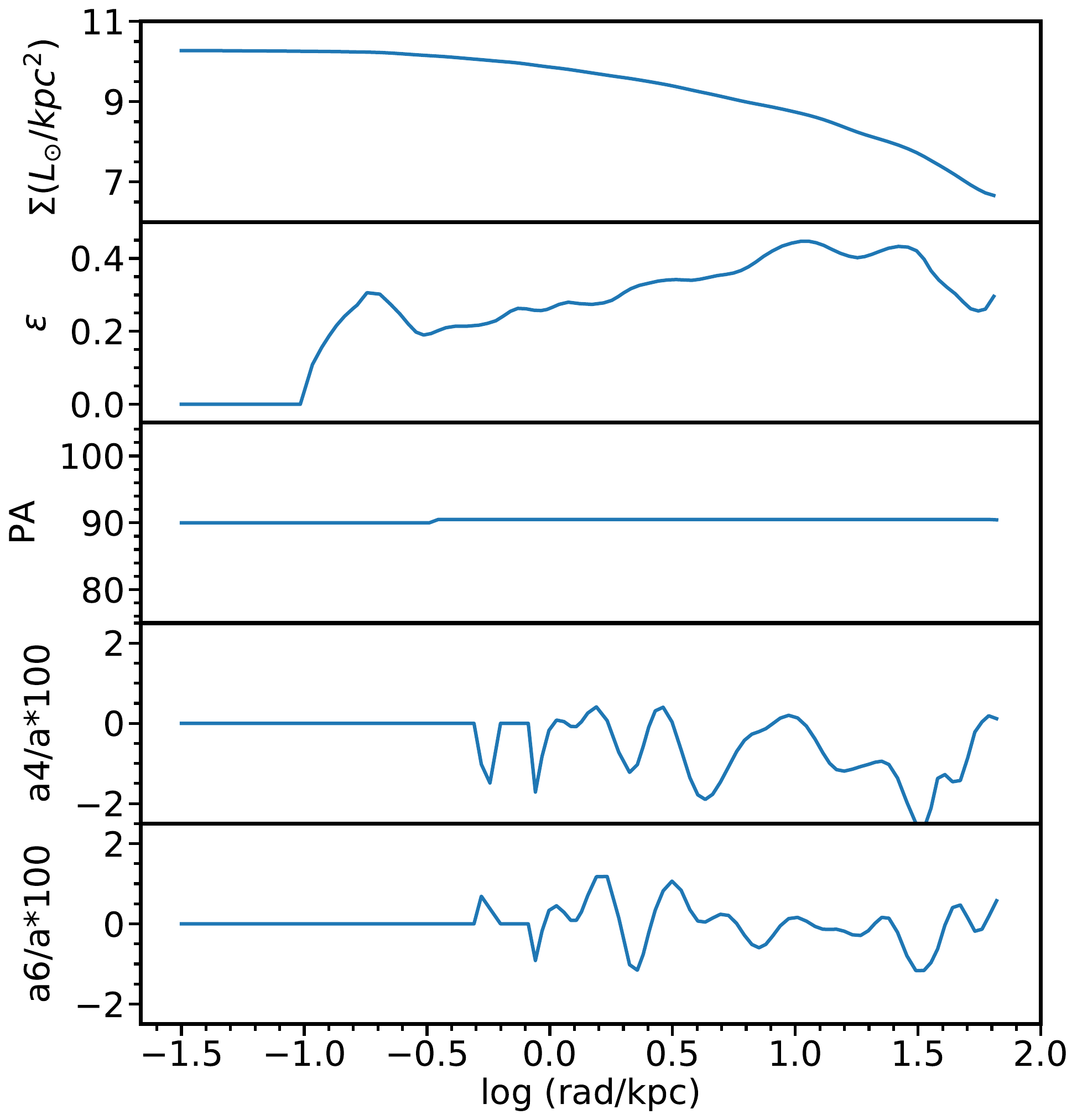}}
\subfloat[MINOR\label{Fig.Isophotes_minor}]{\includegraphics[width=.35\linewidth]{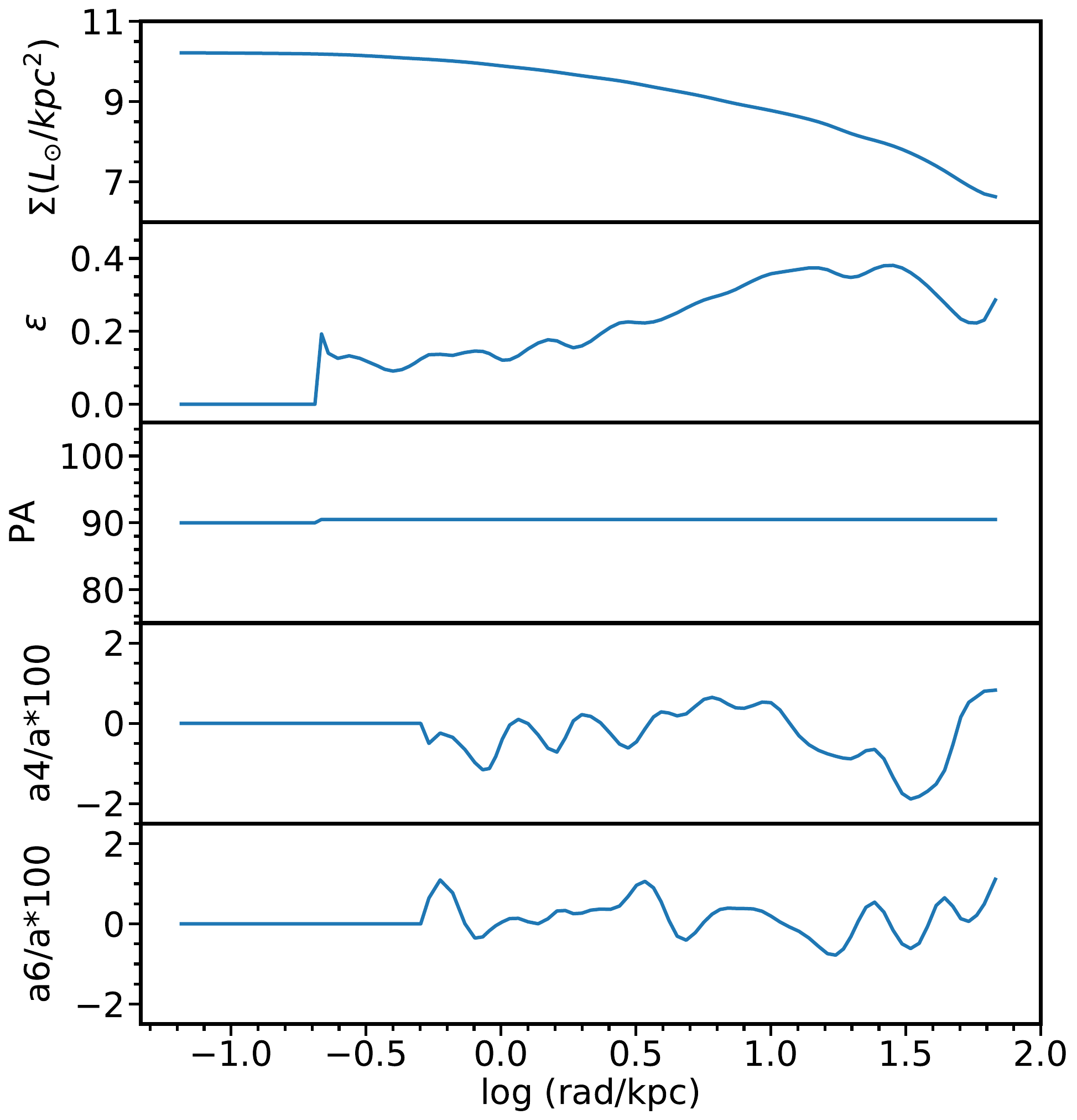}}

    \caption{Isophotes of the four projections (Tab.~\ref{Tab.projections}) of the $N$-body simulation considered throughout this paper. For each projection, the four panels show (from top to bottom) the logarithmic surface brightness $\Sigma$, the ellipticity $\varepsilon$, the position angle PA, and higher-order distortions a$_4$ and a$_6$ of the isophotes as a function of the semi-major axis length.}
    \label{Fig.Isophotes}
\end{figure*}

\begin{figure}
    \centering
    \includegraphics[scale=.2]{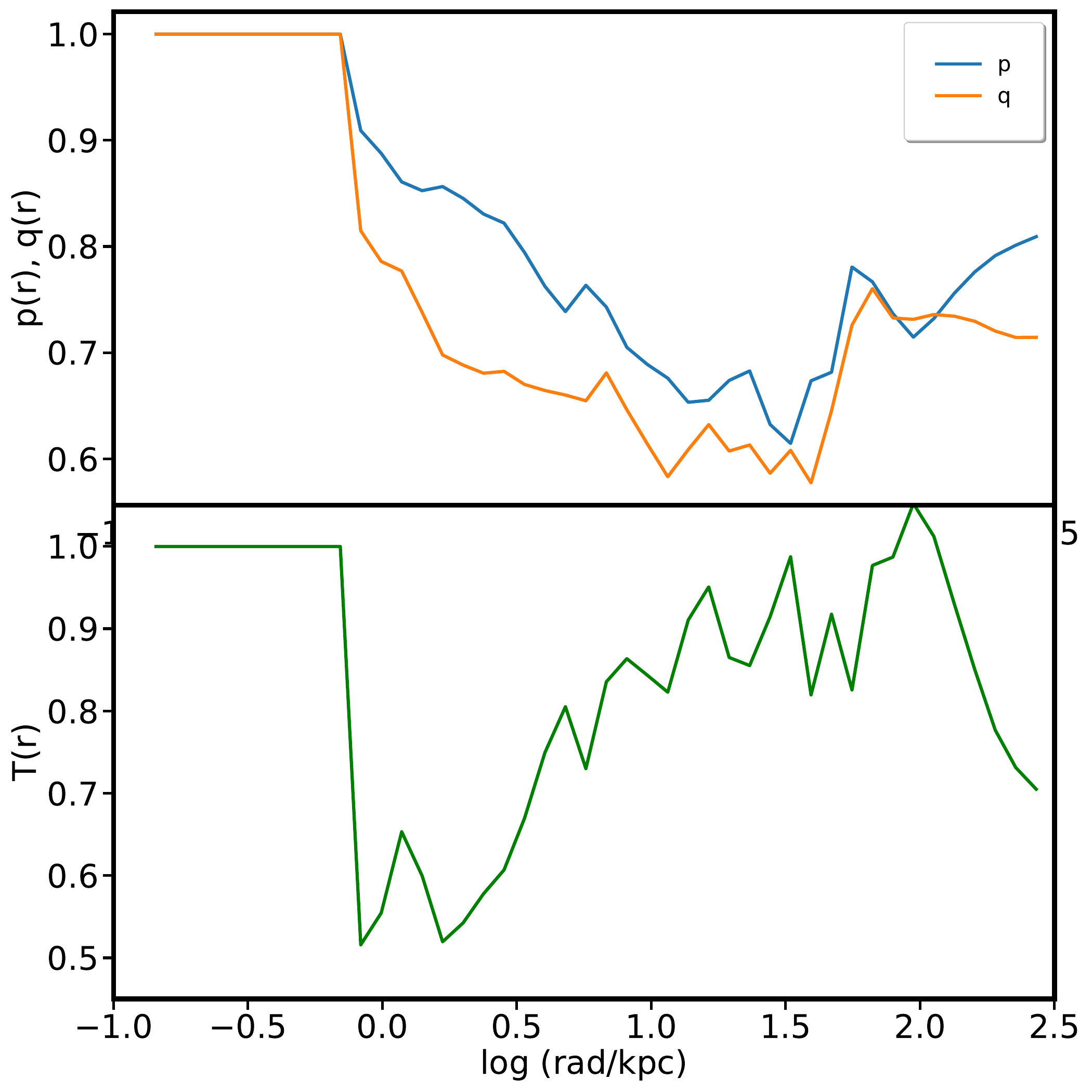}
    \caption{$p(r), q(r)$ (top panel) and $T(r)$ (bottom panel) profiles derived for the $N$-body simulation. The galaxy is spherical in the central regions, reaches the maximum triaxiality at $\sim$3 kpc and then becomes prolate in the outskirts.}
    \label{Fig.pq_true}
\end{figure}

\begin{figure}
    \centering
    \includegraphics[width=40mm]{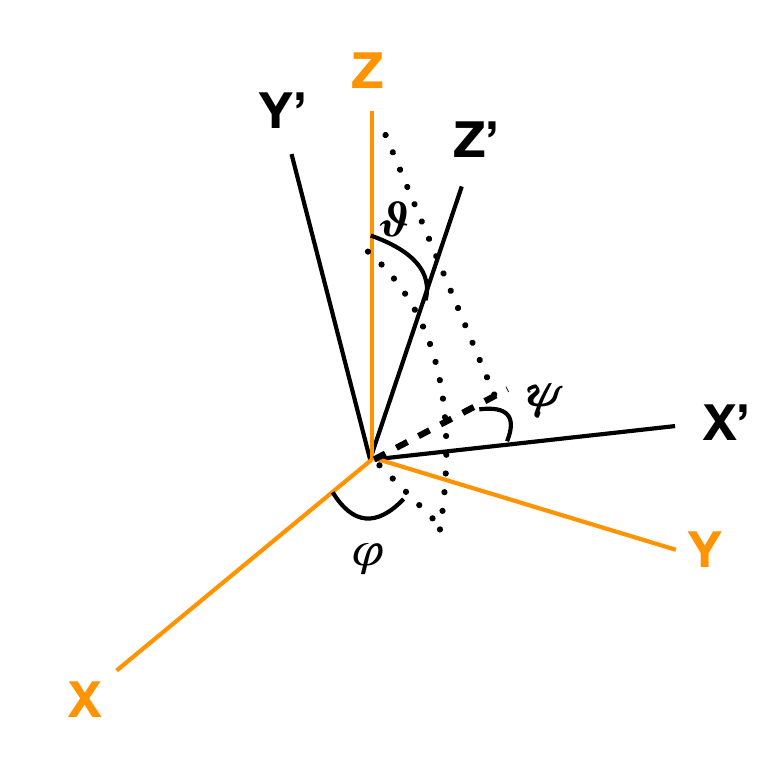}
    \caption{Geometric meaning of the viewing angles $\theta$ and $\phi$, which determine the LOS direction $Z'$, and $\psi$, which is a rotation around the LOS itself.}
    \label{Fig.FOR}
\end{figure}

We test our strategy using an $N$-body simulation extensively described in \citet{Rantala18, Rantala19}. The simulation follows the merging process of two gas-free elliptical galaxies with supermassive black holes and was originally performed to study the formation and the evolution of the so-called core elliptical galaxies, whose light-deficient central regions are thought to be generated by (multiple) BH scouring events \citep{Faber97, Merrit06, Kormendy09, Jens16, Rantala18, Rantala19, Kianusch19}. \\
The $N$-body simulation shows features commonly observed in massive early-type galaxies and closely resembles NGC1600 \citep{Jens16}. Projected on the sky, it is relatively round in the central regions and becomes more flattened when moving out to large radii (see Fig.~\ref{Fig.Isophotes}). In terms of the internal shapes, the galaxy is close to spherical in the central core and then becomes triaxial at large radii (see Fig.~\ref{Fig.pq_true}). The anisotropy $\beta$, defined as
    
       \begin{equation}
           \beta = 1 - \frac{\sigma^2_\theta + \sigma^2_\phi}{2 \sigma^2_r},
       \end{equation}
       
 is negative in the central regions. Here $\sigma_r$, $\sigma_\theta$, $\sigma_\phi$ are the components of the velocity dispersion in spherical coordinates. This tangential bias is observed in several massive ellipticals \citep{Jens14, Jens16, Kianusch19}. It is likely due to the core scouring mechanism, which leads to an ejection of stars with radial orbits in the central regions \citep{Rantala18, Rantala19, Frigo21}. \\
We tested four different projections of the simulation at four different viewing angles (see Fig.\tild\ref{Fig.FOR}). Two projections are along the intermediate (INTERM) and the minor (MINOR) axis of the galaxy\footnote{We do not discuss the projection along the major axis, see Paper I.}, respectively, one lying exactly in between (MIDDLE) and finally another one at random viewing angles (RAND). These
are summarized in Tab.~\ref{Tab.projections}, while their isophotes are shown in Fig.~\ref{Fig.Isophotes}. 
The projections along the principal axes are the only ones where the
deprojection is degenerate even if the surface brightness profile would \textit{exactly} be the projection of a "deformed ellipsoid" model (see Sec.\tild\ref{Sec.depro} below), since
at least one of the two intrinsic shape parameters $p(x)$ or $q(x)$ is not constrained when the LOS coincides with one of the principal axes of the galaxy. MIDDLE is a case we
already considered in dN20, while RAND happens to be the only projection showing a significant isophote twist $\left(\sim 20^\circ\right)$, although the isophotes only show a weak ellipticity $\varepsilon \leq 0.15$.

\subsection{Deprojection} \label{Sec.depro}
Our triaxial deprojection algorithm SHAPE3D is extensively described in
dN20. Here we just report its main features for the sake of the reader's
convenience.
\begin{itemize}
    \item The surface brightness and $\rho$ are placed onto polar elliptical and ellipsoidal
grids, respectively;
    \item The algorithm works under the assumption that a galaxy can
be described by what we call a "deformed ellipsoid", namely an
ellipsoid whose radius is given by
    \begin{equation}
        m^{2-\xi(x)} = x^{2-\xi(x)} + \left[\frac{y}{p(x)}\right]^{2-\xi(x)} + \left[\frac{z}{q(x)}\right]^{2-\xi(x)}
        \label{eq.def_ellips}
    \end{equation}

\noindent where the exponent $\xi$ can be used to generate disky ($\xi > 0$) or
boxy ($\xi < 0$) iso-density surfaces. The three one-dimensional functions $p(x), q(x)$,
and $\xi(x)$, along with the density on the x-axis $\rho_x (x)$, specify $\rho$ at each point of the grid.
    \item Since the algorithm is not a fully parametric method, it requires regularization. What the code minimizes is the likelihood $L = -\frac{\chi^2}{2} + P$
where $\chi^2$ compares the differences between
the observed and the modeled surface brightness and $P$ is a penalty function used
to disfavour unsmooth solutions.
\end{itemize}

\begin{table*}
   
   \begin{tabular}{c c c}
Projection & $\left(\theta,\phi,\psi\right)^\circ$ & Remarks \\   
\hline \hline
MIDDLE & (45,45,45)$^\circ$ & - \\ 
RAND & (60.4,162.3,7.5)$^\circ$ & Viewing angles have been drawn randomly \\ 
INTERM & (90,90,90)$^\circ$ & Projection along y. $p(r)$ photometrically unconstrained \\ 
MINOR & (0,90,90)$^\circ$ & Projection along z. $q(r)$ photometrically unconstrained \\ 
\hline
\end{tabular}
    
\caption{The four projections of the $N$-body simulation studied in this paper. \textit{Col 1:} The projection name. \textit{Col 2:} The projection angles. \textit{Col 3:} Notes on the individual projection.} 
\label{Tab.projections}

\end{table*}

\subsection{Dynamical Modeling}

SMART, the code that we use to compute dynamical models is the triaxial sibling
of the Schwarzschild axisymmetric routine of \citet{Jens04}. The code was presented in BN21. Schwarzschild models are very flexible and do not require \textit{a priori} assumptions on the shape
and/or the anisotropy of the galaxy. As done above for the deprojection
routine, we report here the most important features of the
code.
\begin{enumerate}
    \item The triaxial 
densities $\rho$ derived from the deprojection of the surface brightness  for a wide range of orientation parameters are considered;
    \item A trial total density is constructed as:
        \begin{equation}
        \rho_{\text{TOT}} = \mbhm \times \delta(r) + \mlm \times \rho + \rho_{\text{DM}}
            \label{eq.SMART}
        \end{equation}

\noindent where the first term is the point-like Keplerian potential coming
from the black hole, the second term yields the stars' contribution
through\,\ml\,and the deprojected density $\rho$ and the third term is the DM density, which we compute assuming a modified gNFW \citep{NFW97} profile (see below);
    \item Poisson's equation is solved to obtain the potential which allows the computation of a representative time-averaged orbit library;
    \item The orbital weights are computed such that the difference
between the modeled and the observed LOSVDs is minimized for the assumed orientation and mass profile, with the stellar density as a Lagrangian constraint.
\end{enumerate}

Steps (ii)-(iv) are then repeated for different \mbh, \ml, orientation parameters and dark halo parameters to find the best model. Like our axisymmetric Schwarzschild code, SMART uses the entire LOSVD as a constraint for the orbit model rather than just Gauss-Hermite moments. Like most Schwarzschild codes we use a penalised maximum-likelihood approach to deal with the large number of formal model variables. Specifically, our code uses a maximum-entropy technique and maximises 
\begin{equation}
   \hat{S} = S - \alpha \chi^2,
    \label{eq.Shannon}
\end{equation}
where $S$ is an entropy function. For the current study we use an entropy term related to the Shannon entropy (see Paper I). The strength of the entropy penalty is controlled by the regularisation parameter $\alpha$ and $\chi^2$ compares the observed LOSVDs to the fitted ones. The optimal choice of the smoothing strength in penalised models is important to avoid both overfitting and oversmoothing.
\citet{Jens22} have derived a generalised information criterion AIC$_p$ for penalised models and demonstrated how it can be used to optimise smoothing strengths in a purely data-driven way. To find the correct value of $\alpha$ that prevents the code from finding solutions which fit the data well but are too noisy or others which are too entropy-biased we minimise
\begin{equation}
   \mathrm{AIC}_p = \chi^2 + 2 \text{m}_\text{eff}.
    \label{eq.AIC}
\end{equation}
It generalises the classical Akaike Information Criterion (AIC) by using the concept of effective free parameters ($m_\mathrm{eff}$, c.f. \citealt{Mathias21}) rather than a count of the number of variables\footnote{$m_\mathrm{eff}$ is calculated by running bootstrap simulations and evaluating the covariance between different noise patterns added to the best-fit model and the response of the fit to this noise.}.
The connection between the smoothing $\alpha$ and $\text{m}_\text{eff}$ can be intuitively understood since the higher the smoothing, the less flexible the model will be, thus resulting in a smaller number of effective free parameters. A new feature of our code is that we individually optimise the smoothing for each trial mass model (see Paper I). 

\citet{Mathias21} have shown that determining the best-fit model for a
galaxy is a model selection problem rather than a classical
parameter-estimation problem. This is due to the number of effective
parameters $m_\mathrm{eff}$ that does not only vary for different
smoothings but also from one mass model to another
\citep{Mathias21}. Hence we also use model selection based on AIC$_p$
to determine the best-fit mass and orientation parameters (see Paper
I).

\section{Methodology}

Our strategy is similar for the four $N$-body projections highlighted
in Tab.\tild\ref{Tab.projections}. We simulate realistic observational
conditions by combining both the deprojection and the dynamical model
as we do for observations of real galaxies. We first run several
deprojections in one octant to shrink the region of possible viewing angles
(Sec.\tild\ref{Ssec.photometry}). Then, we dynamically model all
plausible densities (Sec.\tild\ref{Ssec.dyn}) in order to find the
viewing angles that give the best agreement with the observed
kinematics and analyze the resulting shapes and anisotropy profiles.

\subsection{Reducing the number of viewing direction with
photometry only} \label{Ssec.photometry}

We want to reduce of the number of viewing angles compatible with
a specific photometric data set. In dN20, we showed that SHAPE3D can deal with this
task and -- moreover -- can recover the correct intrinsic 3D density $\rho$ (for the correct viewing angles). This is true as long as the object under study can be approximated by a nearly ellipsoidal shape and overcomes the degeneracy that is inherent to deprojections in general (see App.
A of dN20)\footnote{Even for ellipsoidal bodies, the uniqueness of the deprojection is only true if we do not project along one of the
principal axes. In Sec.\tild\ref{Ssec.inclination} we discuss a possible solution for this case.}. Our approach is described in detail below.
\begin{enumerate}
    \item We generate the four galaxy images corresponding to each of the
four $N$-body projections (Tab.\tild\ref{Tab.projections}) by projecting the intrinsic density
calculated from the particles. Differently from what we did in dN20,
here we choose not go through isophotal fits and directly fit the
projected image\footnote{If we had chosen to go through isophotal fits, then the level of noise in the images would be
smoothed out since it would be impossible to reproduce the noisy isophotes properly
using Fourier coefficients.}.
   \item We define a $\left(\theta, \phi, \psi\right)$ grid, with $\left(\theta, \phi\right)$ going from 0$^\circ$ to 90$^\circ$ and $\psi$ going from 0$^\circ$ to 170$^\circ$, each angle with a step of 10$^\circ$. This is justified as long as we can assume triaxial symmetry, i.e. that the galaxy appears identical when viewed from different octants. For the tests in this paper this is guaranteed because we average the particle distributions in different octants (see also eq. 10 and Fig. 1 of dN20). The assumption works also well
   for the most massive ellipticals which do not show significant disky
    features or bars. 
\item  We deproject the surface brightness images for every possible viewing
direction that we have defined on our $\left(\theta, \phi, \psi\right)$ grid. Each deprojection is carried out using a 30 $\times$ 12 grid for the surface brightness and a 50 $\times$ 13 $\times$ 13 grid for $\rho$. The smallest radius of both grids is at 0.05 kpc (for the simulation 1 kpc $\sim$ 10 asec), whereas the largest radii are at 68 kpc for surface brightness and at 270 kpc for $\rho$. 
 \item We select a threshold for the RMS, RMS$_\text{thr}$, above which a viewing direction
is discarded. The values we choose are 0.01 mag
for RAND, 0.013 mag for MIDDLE and 0.02 mag
for MINOR and INTERM. These thresholds allow us to discard $\sim$90\% of the deprojected densities.
\item We then discard all densities showing $p(r)$ \& $q(r)$ profiles such
that the order relation between the principal axes is not conserved
at all radii, i.e. profiles intersecting with each other. We do this because, as shown in dN20, such solutions are likely to generate twists $>$ 40$^\circ$, which are not observed in (relaxed) massive ellipticals. All viewing directions fulfilling this requirement and yielding an RMS smaller than RMS$_\text{thr}$ are dynamically modeled. As one can see from Tab.~\ref{Tab.numbers}, the deprojection cut-off does a very good job in reducing the number of viewing angles compatible with the photometry, given that we typically need to sample only $\sim$5\% of the deprojected light distributions. 
\item Finally, when we look at a galaxy along one of the principal
axes\footnote{A clue about a galaxy orientation along the principal is e.g. the lack of isophotal twist, which cannot occur in this case.} $\rho$ cannot be uniquely constrained by the projected surface brightness alone. In order to test to which extent we can probe the full range of allowed densities, we take the MINOR case as an example and consider various different values for the unconstrained shape parameter (in this case $q(r)$) and run additional deprojections. This is only needed when the LOS lies on one of the principal axes, meaning that we need to perform this exercise at $\left(\theta, \phi\right) = \left(0, 90\right)^\circ$. For each one of these light densities, we need to launch a separate set of dynamical models keeping the viewing angles fixed but varying the 3D density. For this exercise, we sample $q(r)$ from 0.5 to 0.8.


\end{enumerate}

\begin{figure*}

\subfloat[MIDDLE\label{Fig.45_Rhopq}]{\includegraphics[width=.3\linewidth]{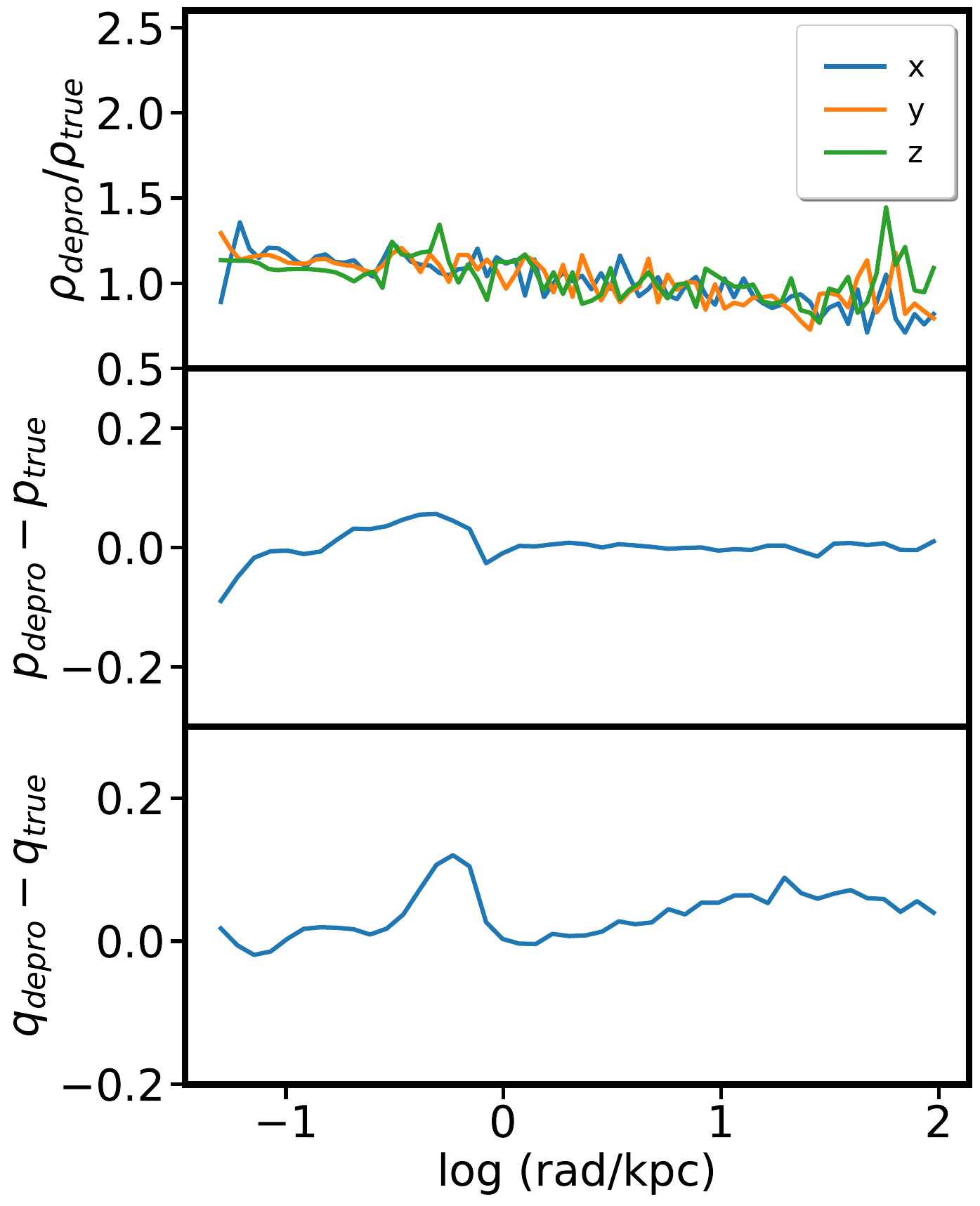}}
\subfloat[RAND\label{Fig.Ran_Rhopq}]{\includegraphics[width=.3\linewidth]{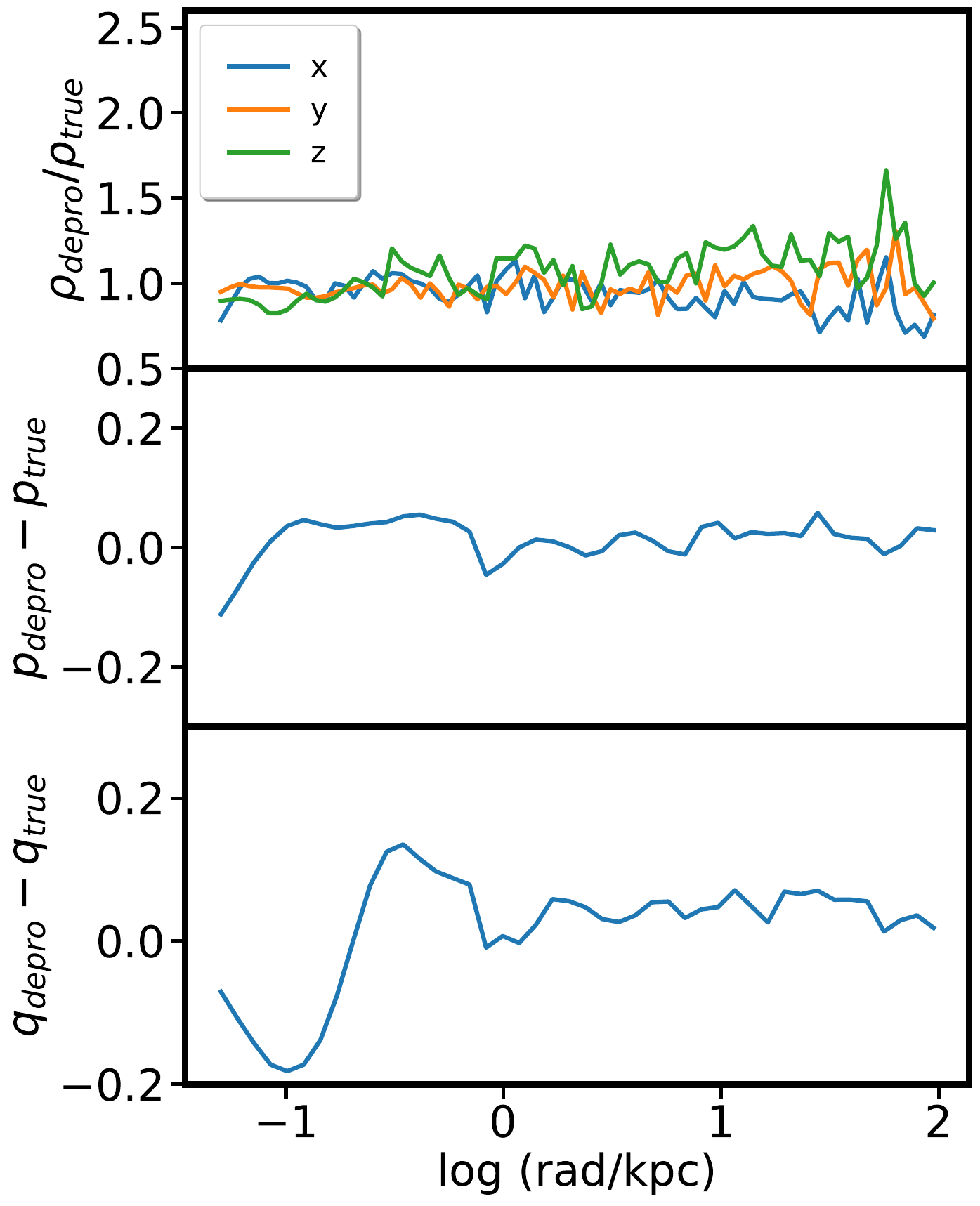}}

\subfloat[INTERM\label{Fig.Int_Rhopq}]{\includegraphics[width=.3\linewidth]{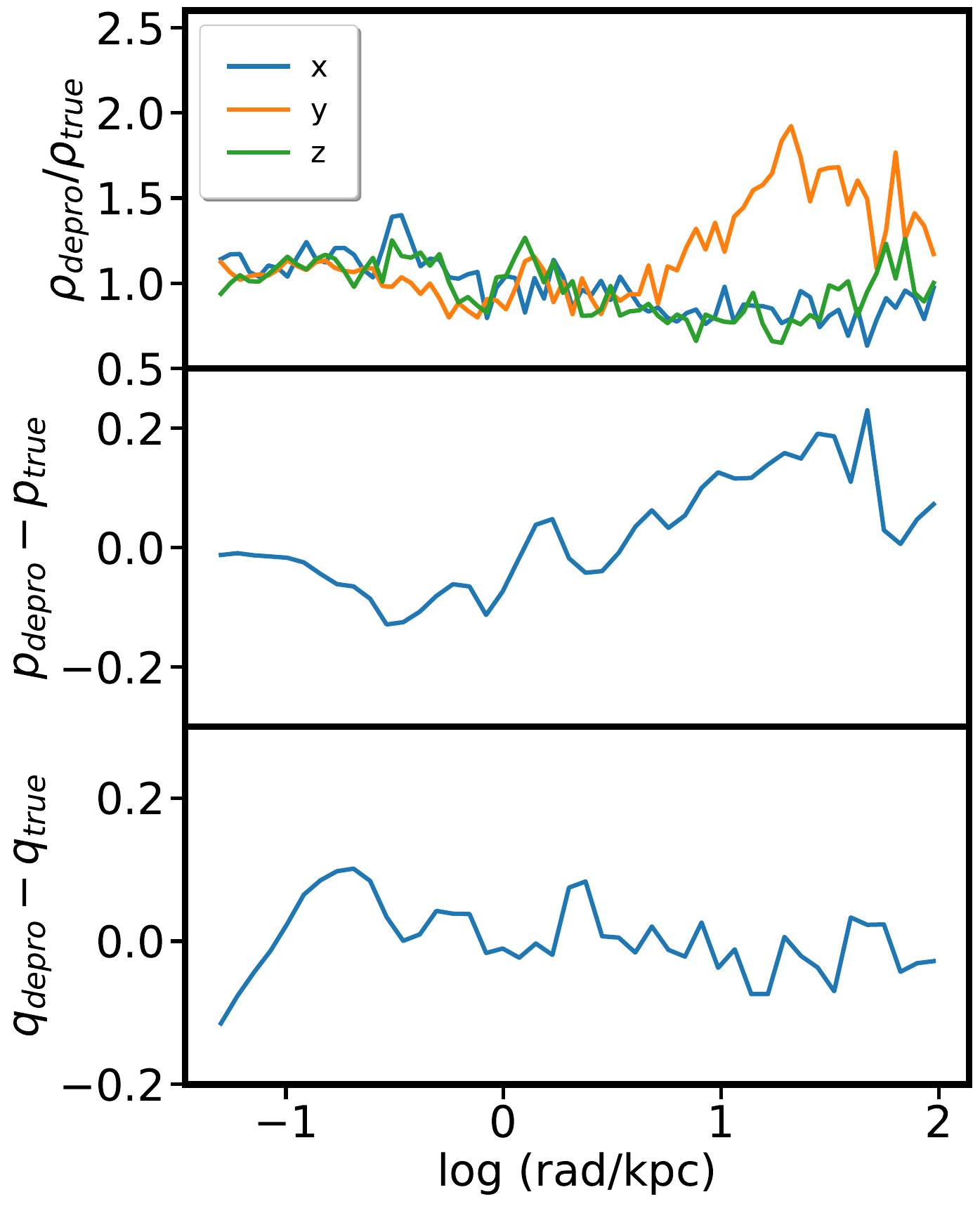}}
\subfloat[MINOR\label{Fig.Min_Rhopq}]{\includegraphics[width=.3\linewidth]{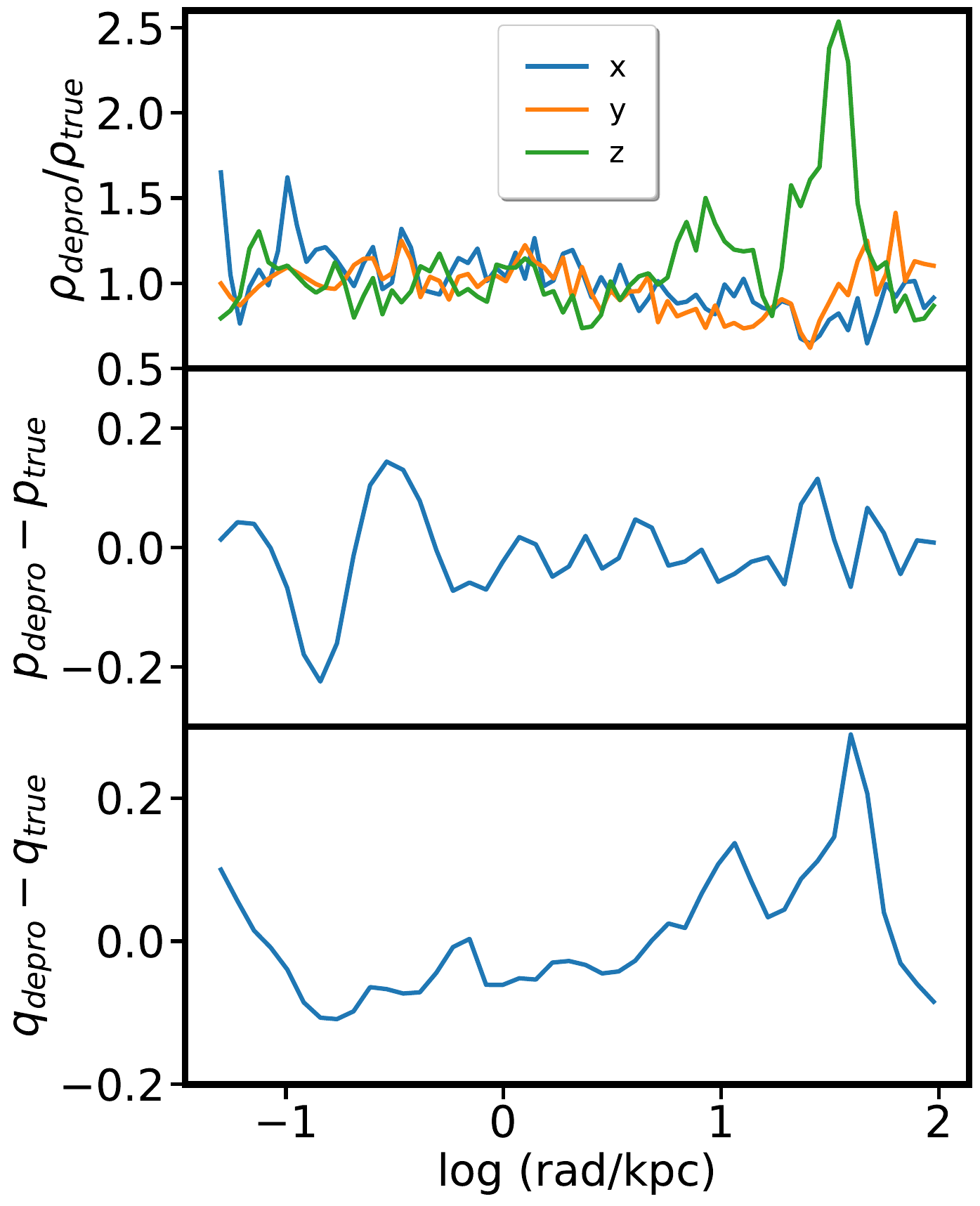}}

    \caption{Comparison between recovered (at the correct viewing angles) and true quantities for the four projections of the $N$-body simulation (Tab~\ref{Tab.projections}) considered throughout the paper. In the top panels we show the ratios of the light densities along the three principal axes, whereas in the bottom panels we show the differences for the corresponding $p(r)$, $q(r)$ profiles.}
    \label{Fig.Rhopq}
\end{figure*}

\begin{figure*}

\subfloat{\includegraphics[width=.35\linewidth]{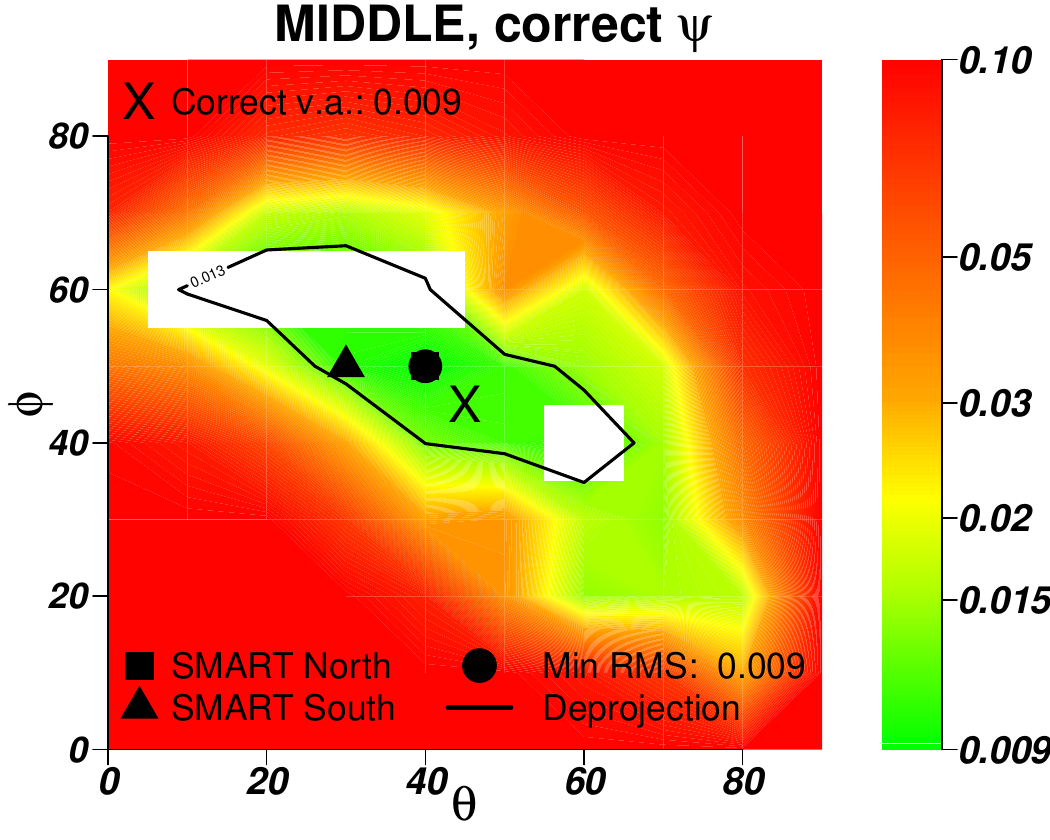}}
\subfloat{\includegraphics[width=.35\linewidth]{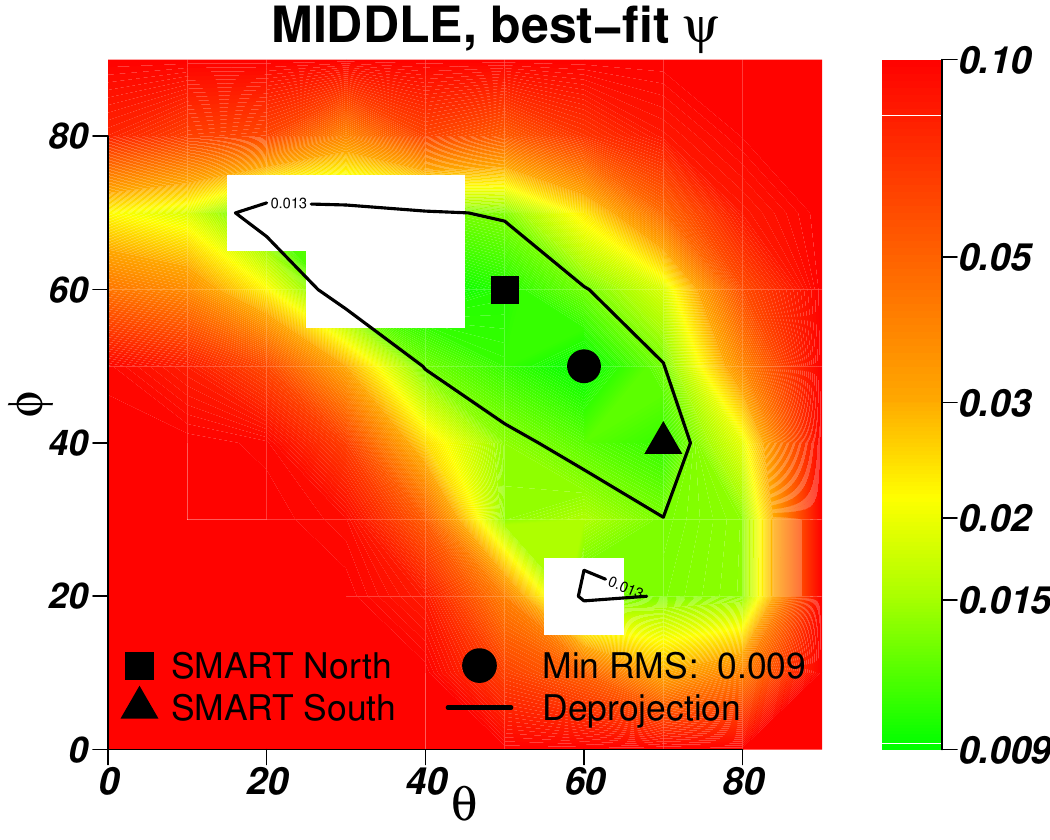}}

\subfloat{\includegraphics[width=.35\linewidth]{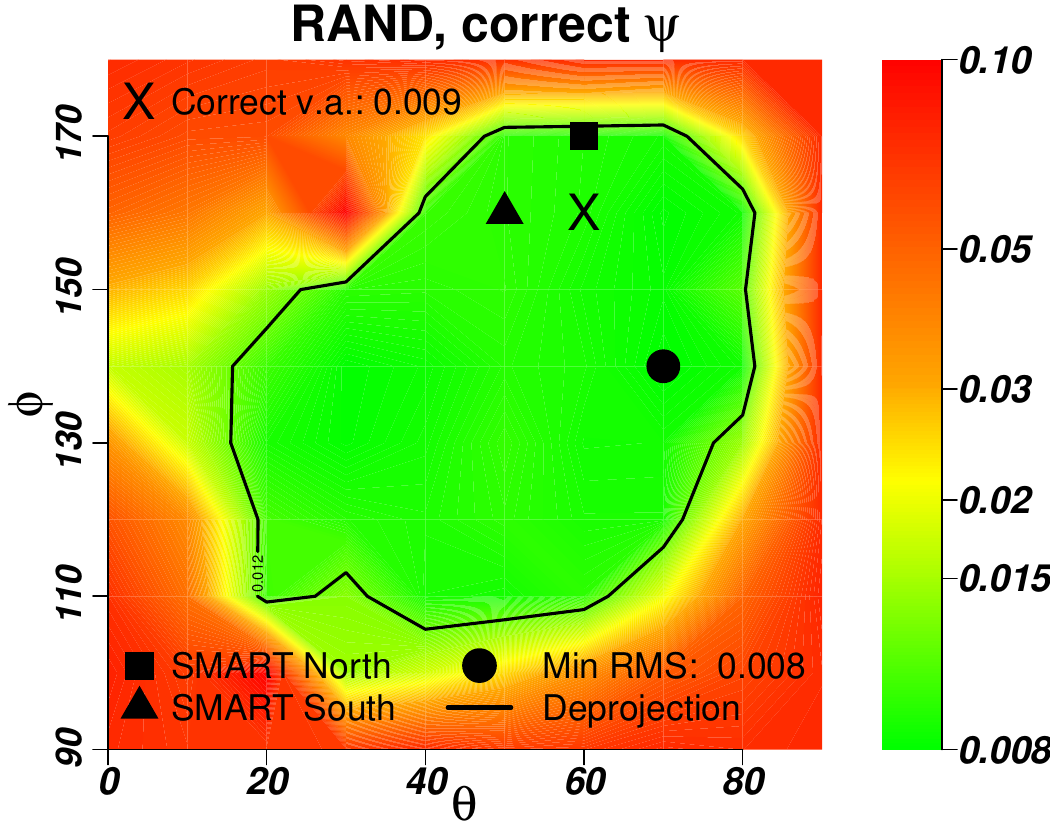}}
\subfloat{\includegraphics[width=.35\linewidth]{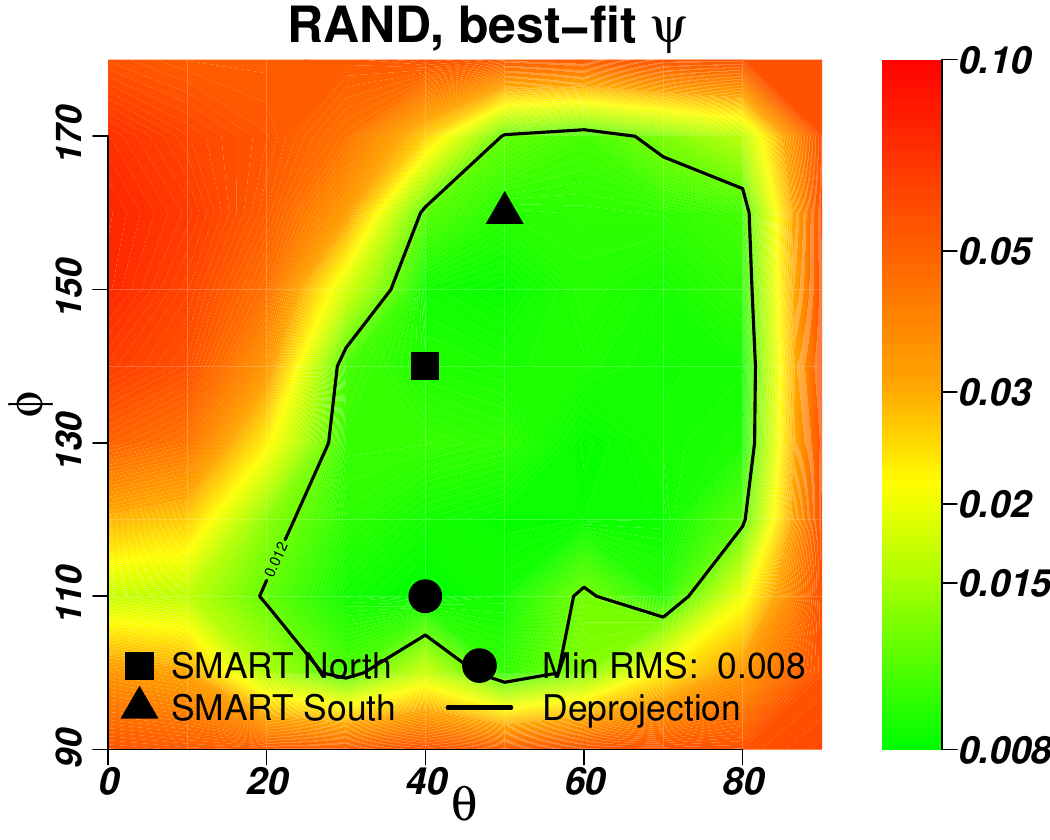}}

\subfloat{\includegraphics[width=.35\linewidth]{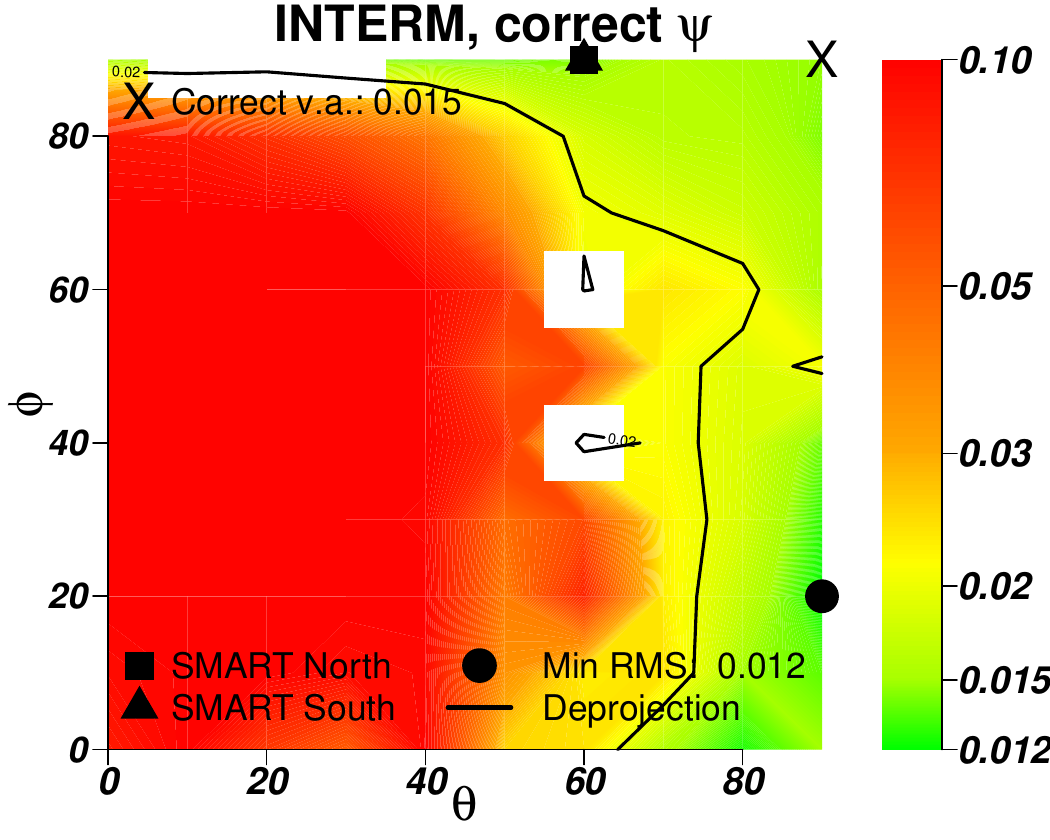}}
\subfloat{\includegraphics[width=.35\linewidth]{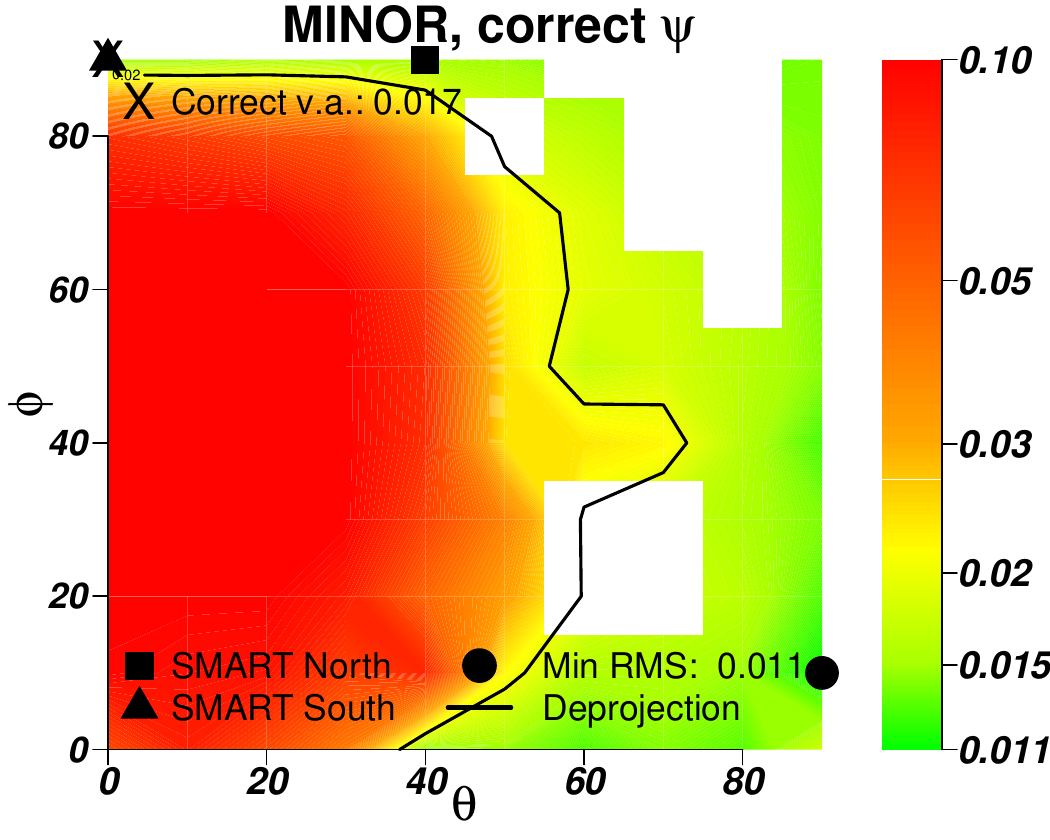}}

    \caption{Logarithmic RMS errors in surface brightness $\Delta \log \Sigma$ (left) for the four $N$-body projections (Tab.\tild\ref{Tab.projections}), obtained for constrained-shape deprojections at different assumed $\theta,\phi$ viewing angles for the correct values of $\psi$ and for the best-fit $\psi$ from the dynamical models. For INTERM and MINOR these two values coincide. The cross labels the correct $\left(\theta,\phi\right)$. This is not shown in the "best-fit" plots because in this case $\psi$ is not the correct one. The black dot is at the best-fit solution from the deprojection, while the black square and triangle indicate the two best-fit solutions from the dynamical modeling. The solid contour delimits the area inside which the RMS in surface brightness is within twice the minimum value we find on the grid. Finally, empty (white) squares depict regions discarded because of crossing $p$ and $q$ profiles.}
    
    \label{Fig.Maps_depro_SMART}
\end{figure*}

\noindent The typical RMS $=\sqrt{\langle \left(\ln ( I_{\text{obs}}/I_{\text{fit}})\right)^2 \rangle}$ between the true and the recovered surface luminosity for the best-fit solutions are $\sim$0.009 for MIDDLE and RAND and $\sim$0.012 for the two projections along the principal axes. \\
From Fig.~\ref{Fig.Maps_depro_SMART} we also see that the correct LOS, specified by $\theta$ and $\phi$, is always included between the densities which we model with SMART. Before moving on to the dynamical model, it is worth making a few comments about the different $N$-body projections:

\begin{itemize}
  \item The RMS$=\sqrt{\langle \left(\ln ( \rho_{\text{true}}/\rho_{\text{fit}})\right)^2 \rangle}$ between the true $\rho$ from the $N$-body simulation is 0.095 and 0.097 for MIDDLE and RAND, respectively. Along the principal axes the RMS is a bit worse ($\sim$0.15), as expected since either $p(r)$ or $q(r)$ cannot be uniquely recovered anymore. For each of the four projections, we show in Fig.~\ref{Fig.Rhopq} the comparison between the observed and the recovered density, as well as between the observed and the recovered $p(r)$ and $q(r)$. The largest deviations are, as expected, observed for the INTERM projection along the $y$-axis (and thus for $p(r)$) as well as the MINOR projection along the $z$-axis (and thus for $q(r)$), since in these cases information about $p(r)$ or $q(r)$ cannot be recovered from the photometry given the position of the LOS.
\item The roundest projection RAND shows the largest number of possible viewing directions (although the threshold is only a factor $\sim$1.1 above the best-fit RMS), which is expected since the isophotes are very round, while the other three projections are
flatter (Fig.\tild\ref{Fig.Isophotes}). On the other hand,
INTERM and MINOR are much better constrained, with
no solutions when $\psi$ is wrong. This is intriguing since their ellipticity profile is similar to the MIDDLE case, for which a much larger variety of solutions is found, \textit{and} with a larger threshold compared to the respective best-fit RMS values. This might be due to the twist being as small as $\sim$0.5$^\circ$, since numeric uncertainties in SHAPE3D might generate solutions with small twists and a profile without twist is particularly simple to fit assuming an inclination along the principal axes. 

\end{itemize}

\begin{figure*}

\subfloat{\includegraphics[width=.3\linewidth]{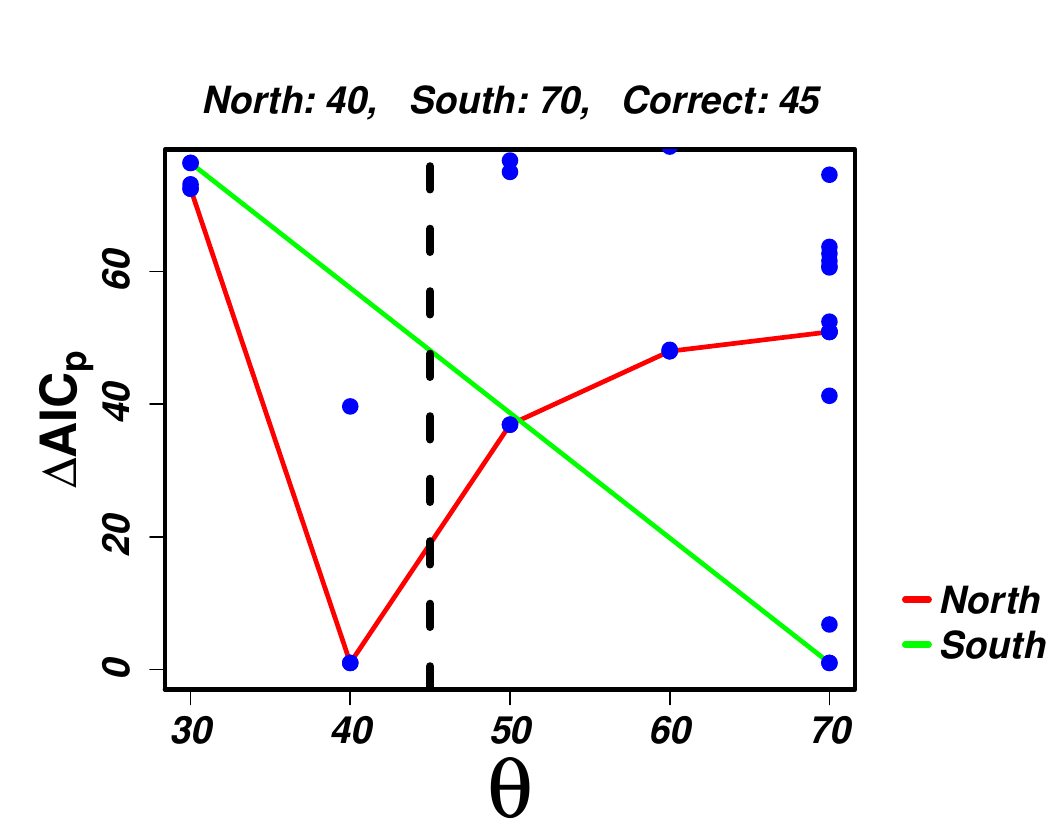}}
\subfloat{\includegraphics[width=.3\linewidth]{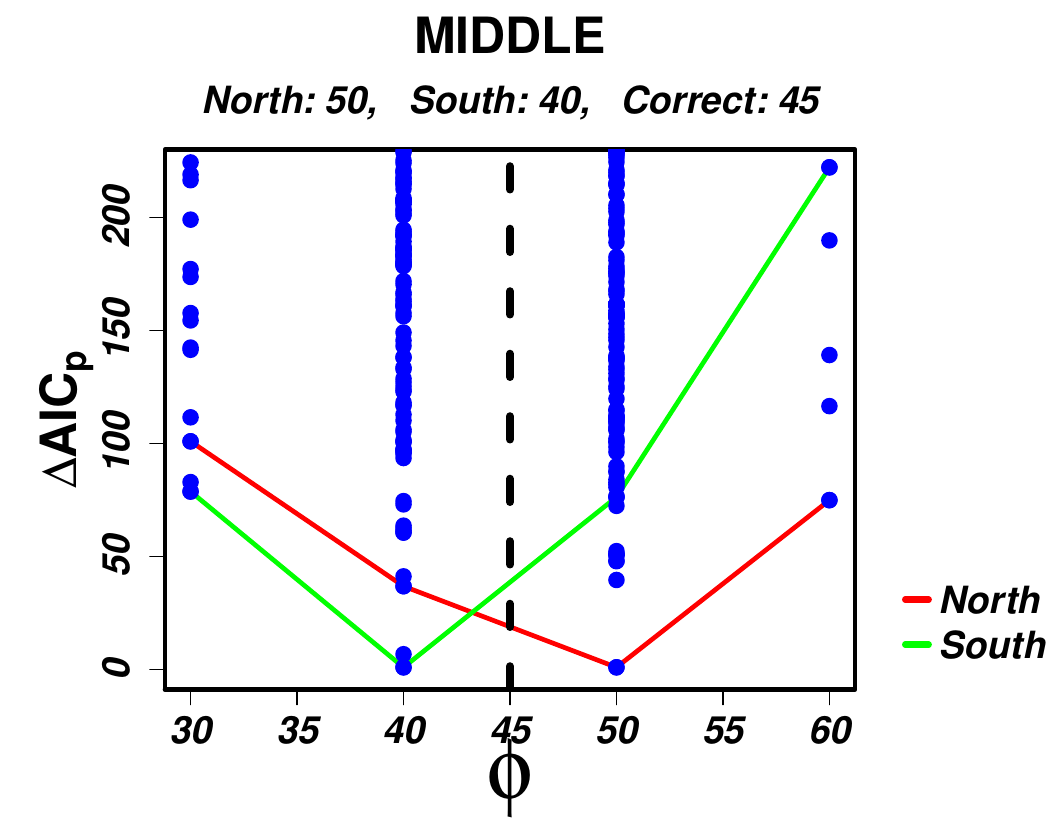}}
\subfloat{\includegraphics[width=.3\linewidth]{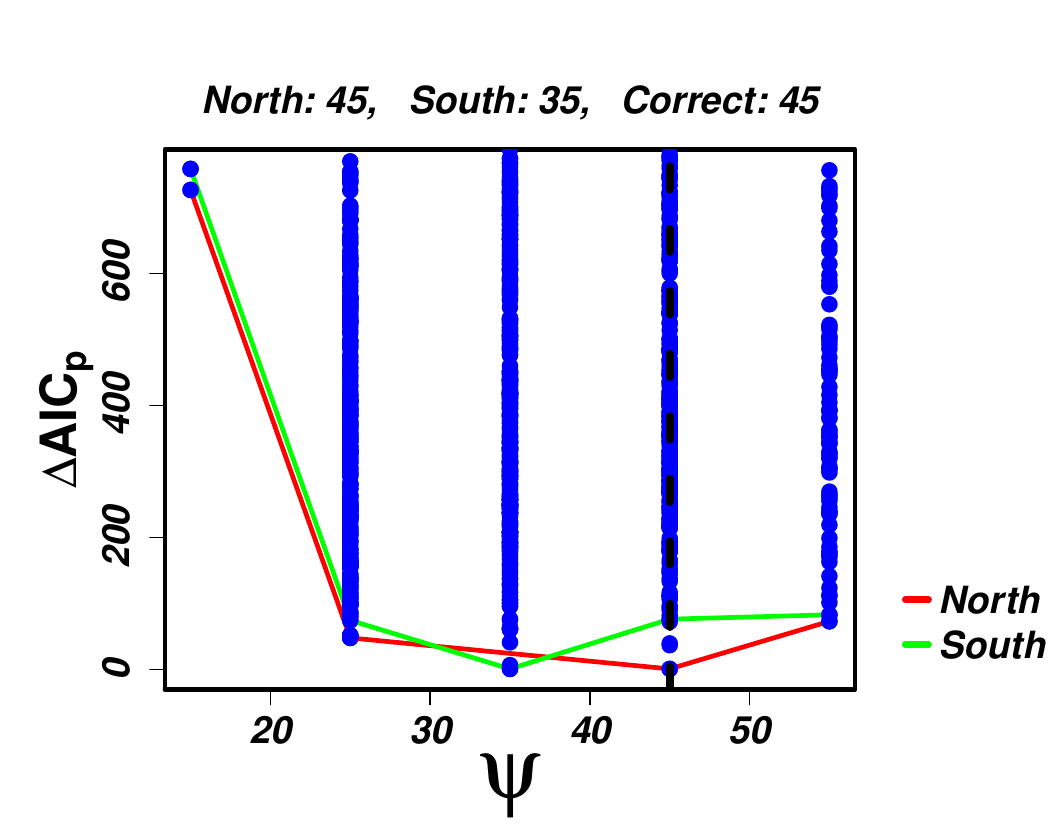}}

\subfloat{\includegraphics[width=.3\linewidth]{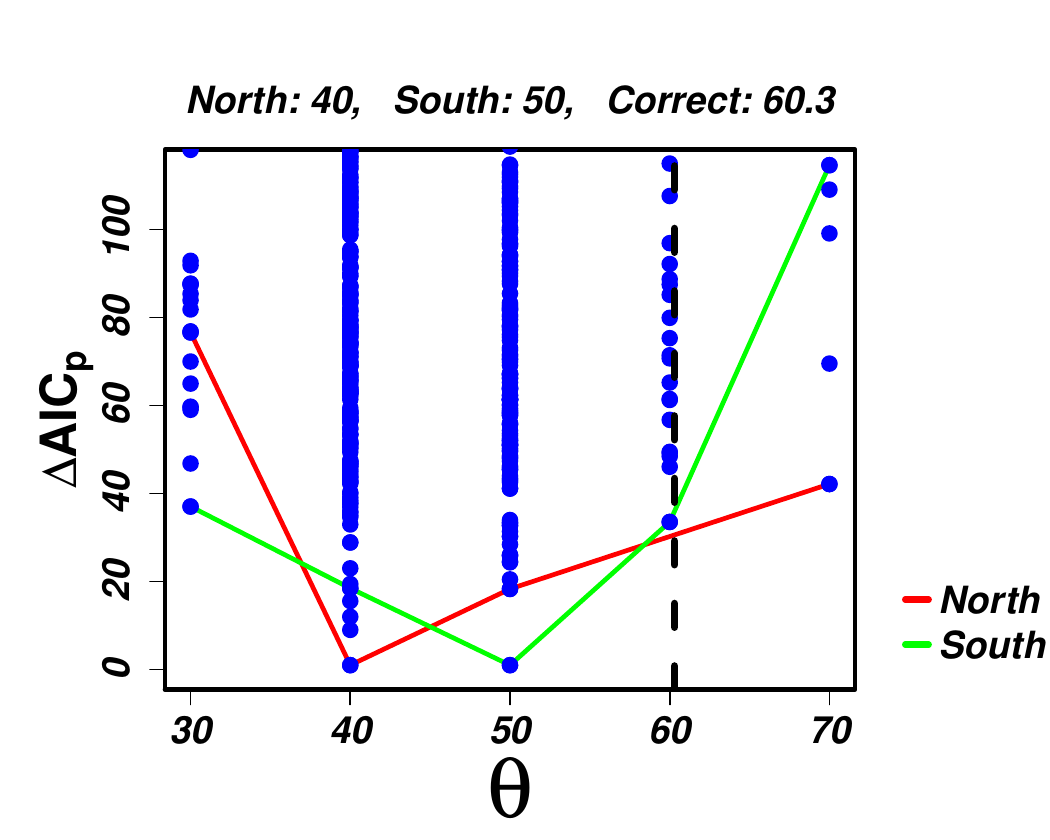}}
\subfloat{\includegraphics[width=.3\linewidth]{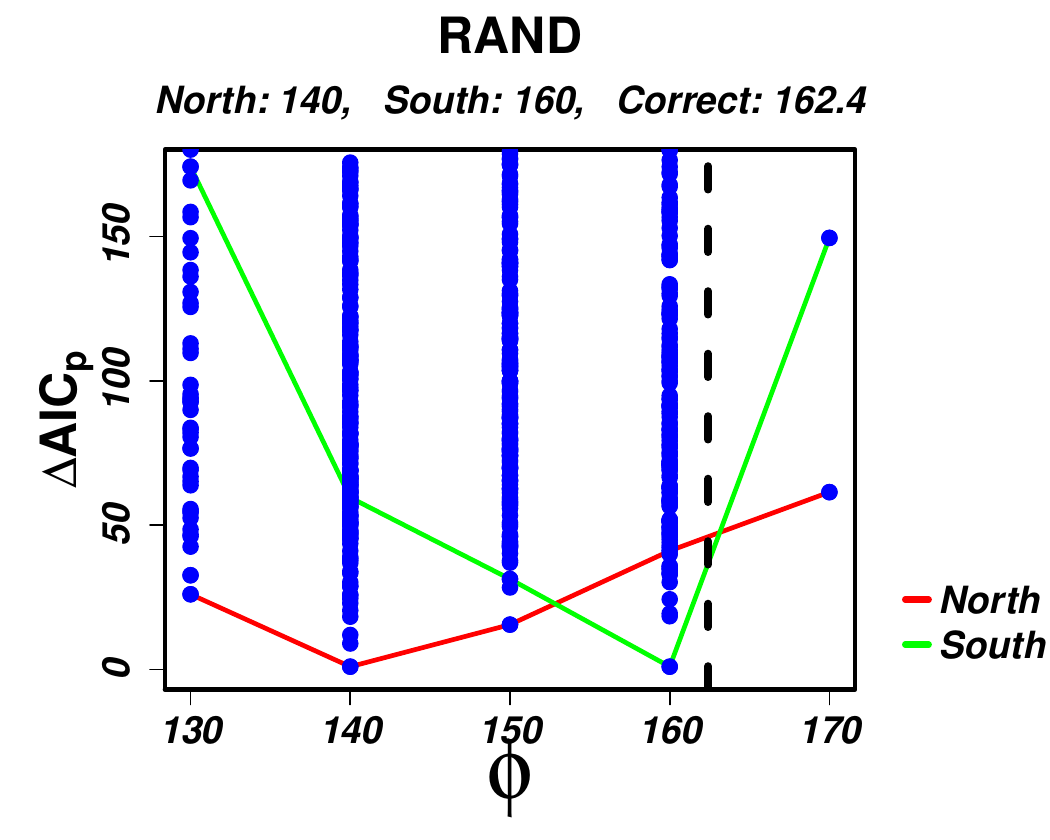}}
\subfloat{\includegraphics[width=.3\linewidth]{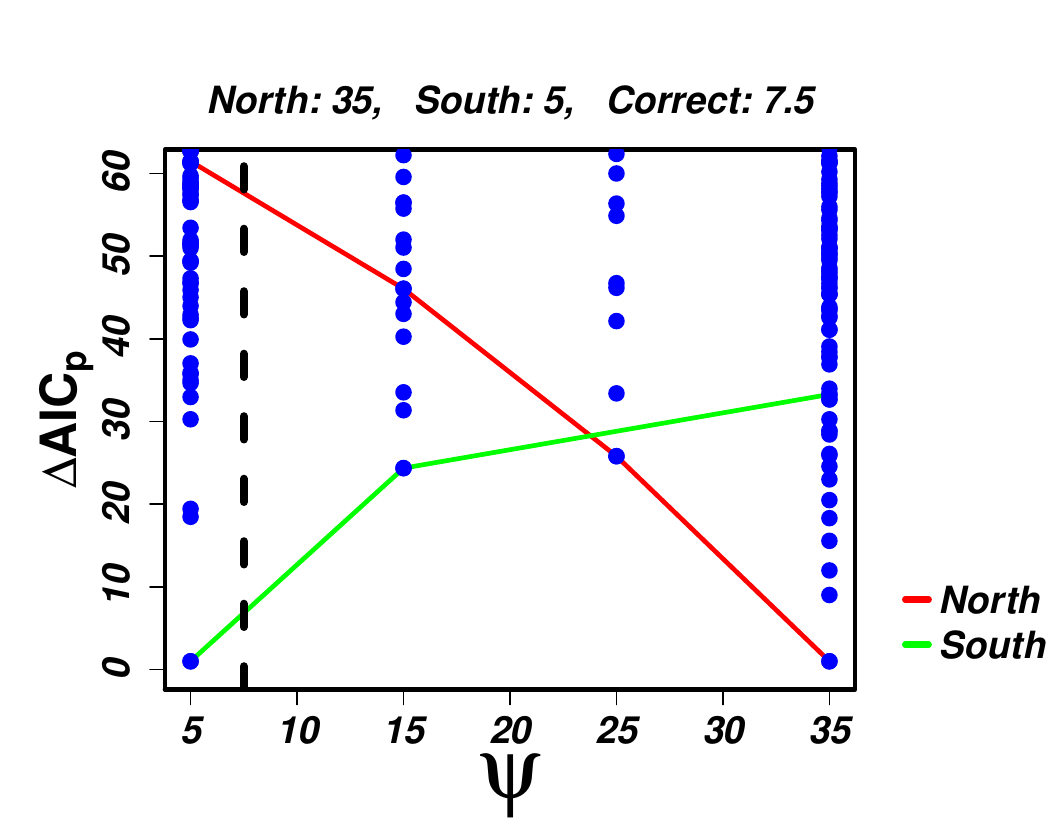}}

\subfloat{\includegraphics[width=.3\linewidth]{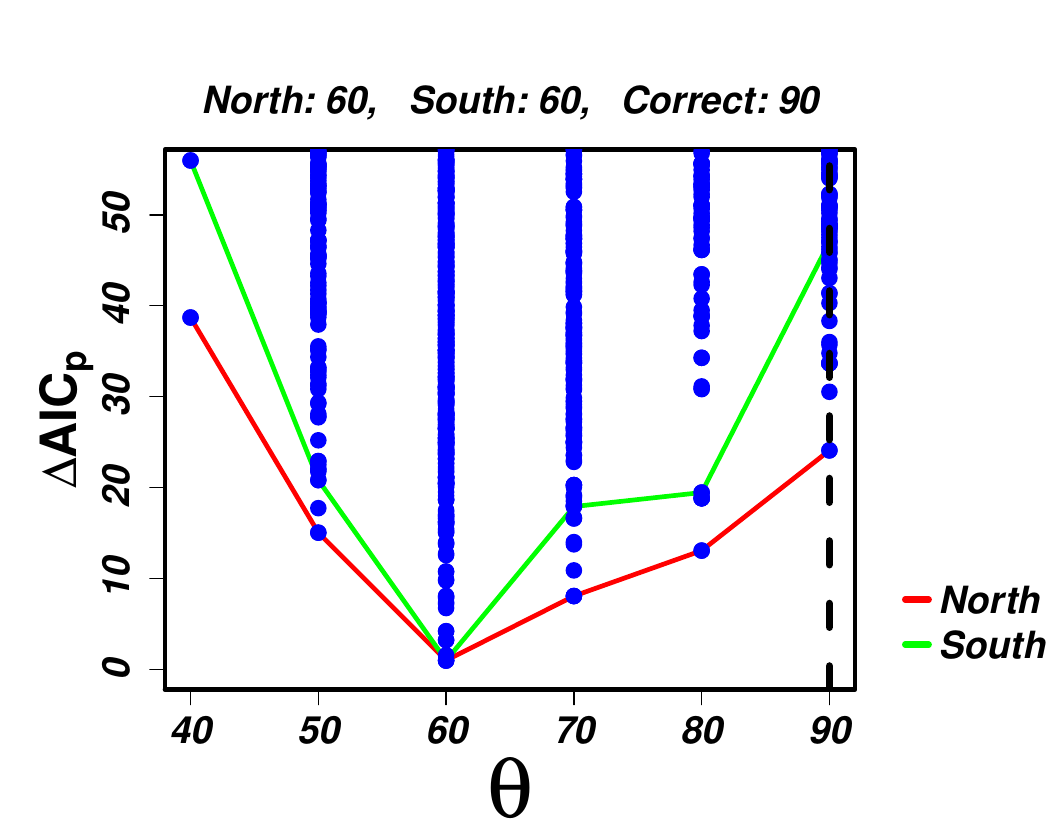}}
\subfloat{\includegraphics[width=.3\linewidth]{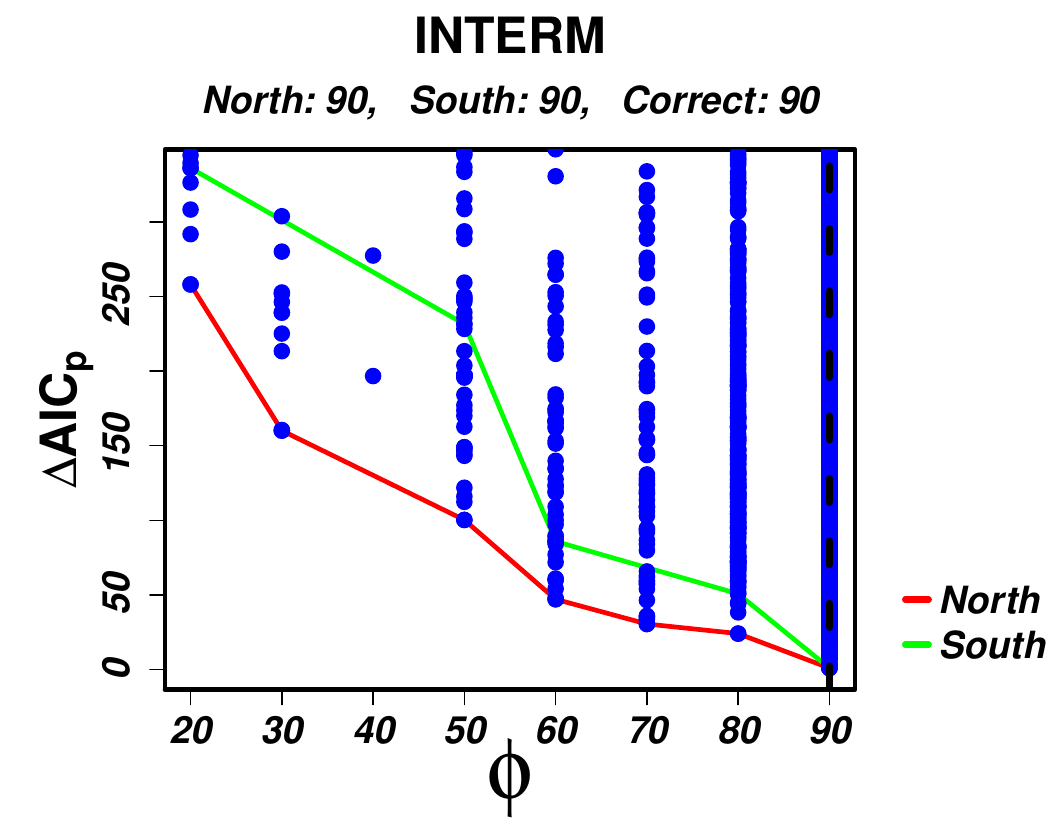}}
\subfloat{\includegraphics[width=.3\linewidth]{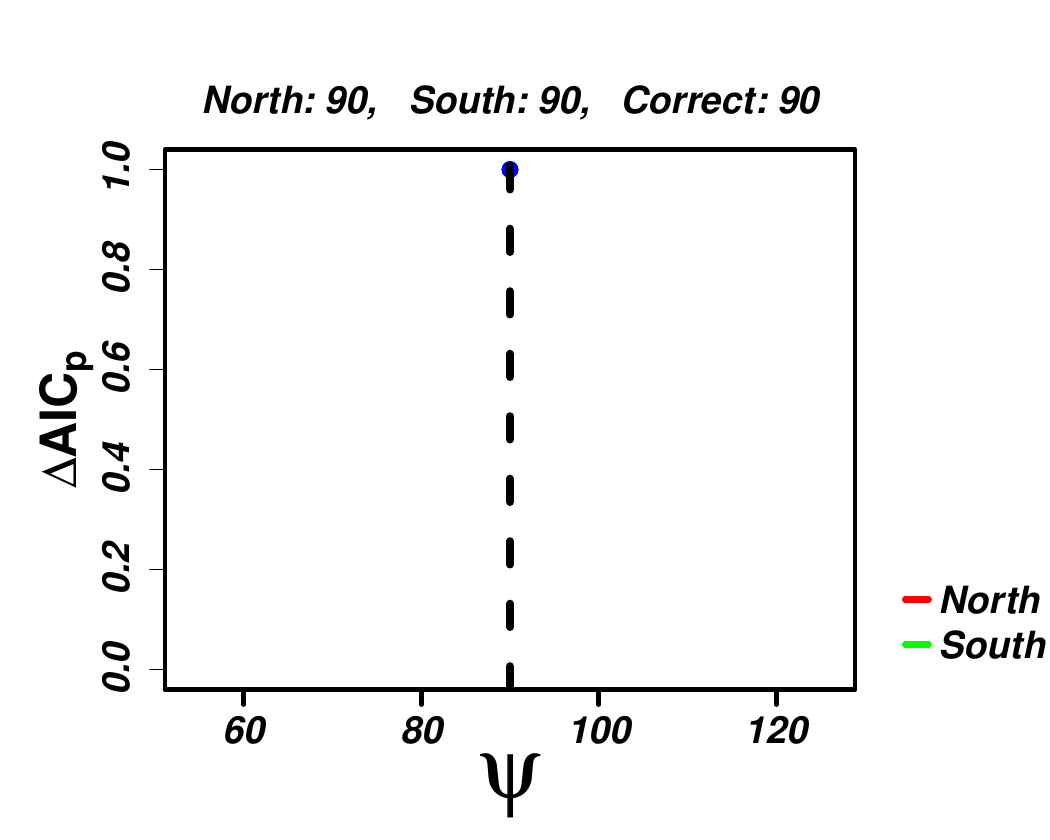}}

\subfloat{\includegraphics[width=.3\linewidth]{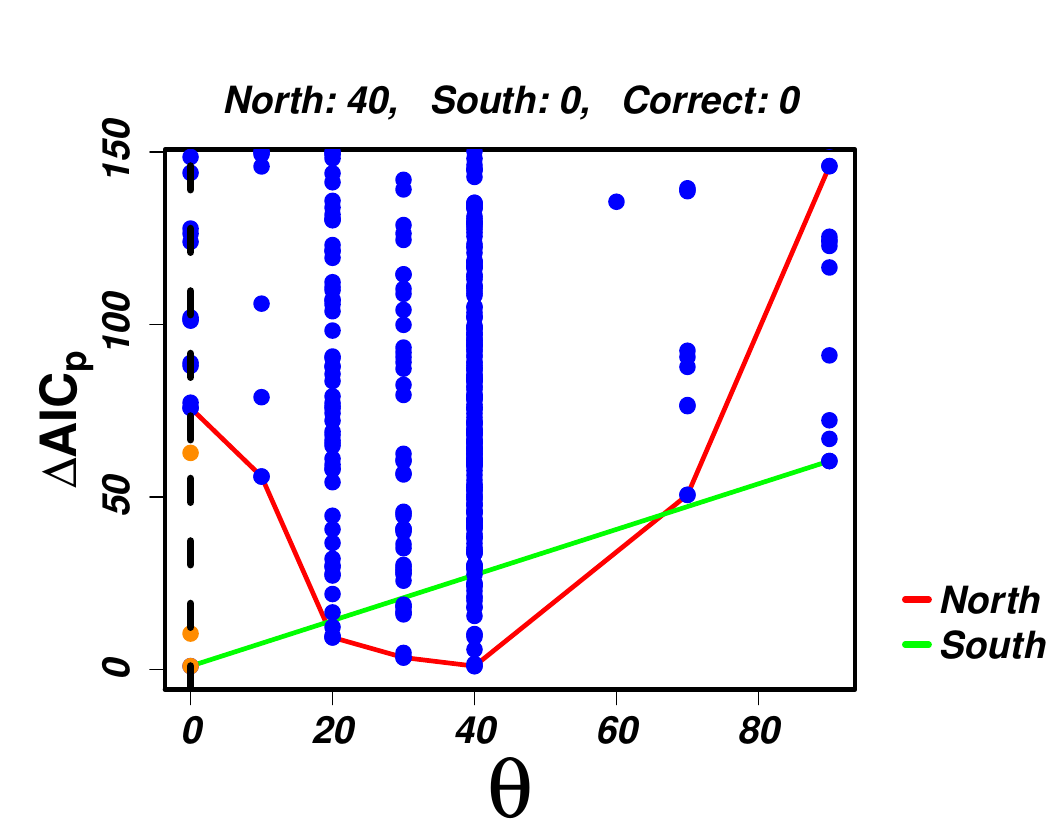}}
\subfloat{\includegraphics[width=.3\linewidth]{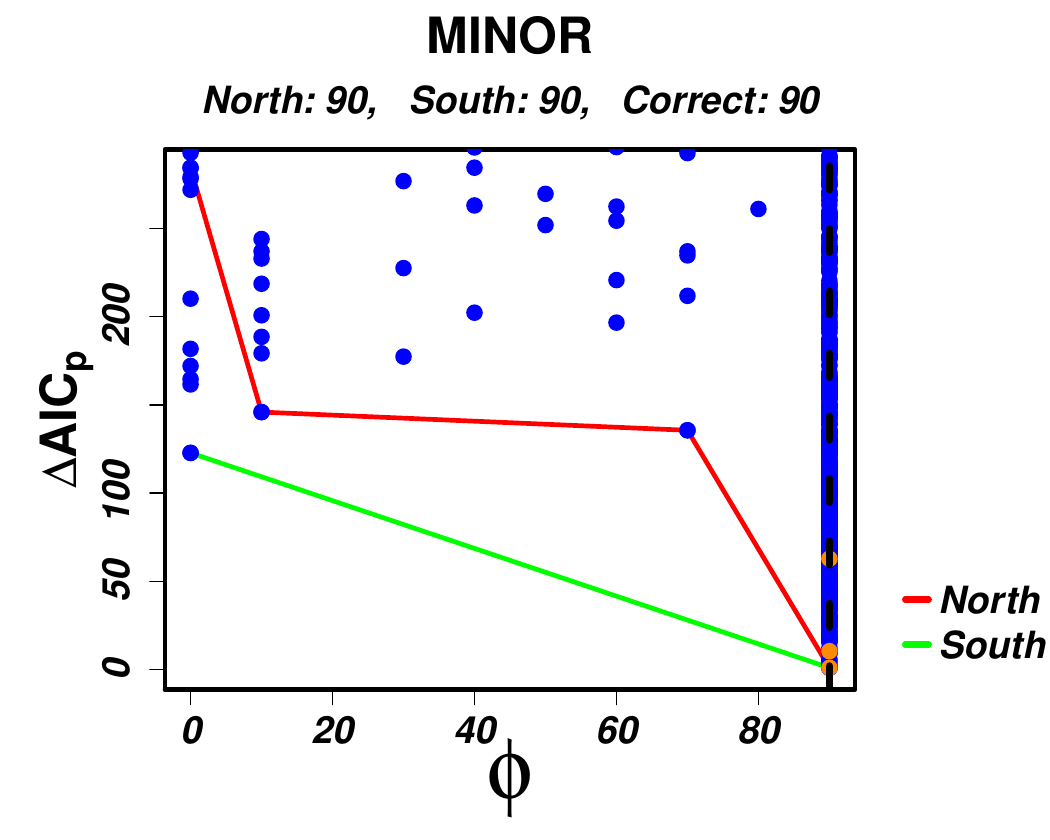}}
\subfloat{\includegraphics[width=.3\linewidth]{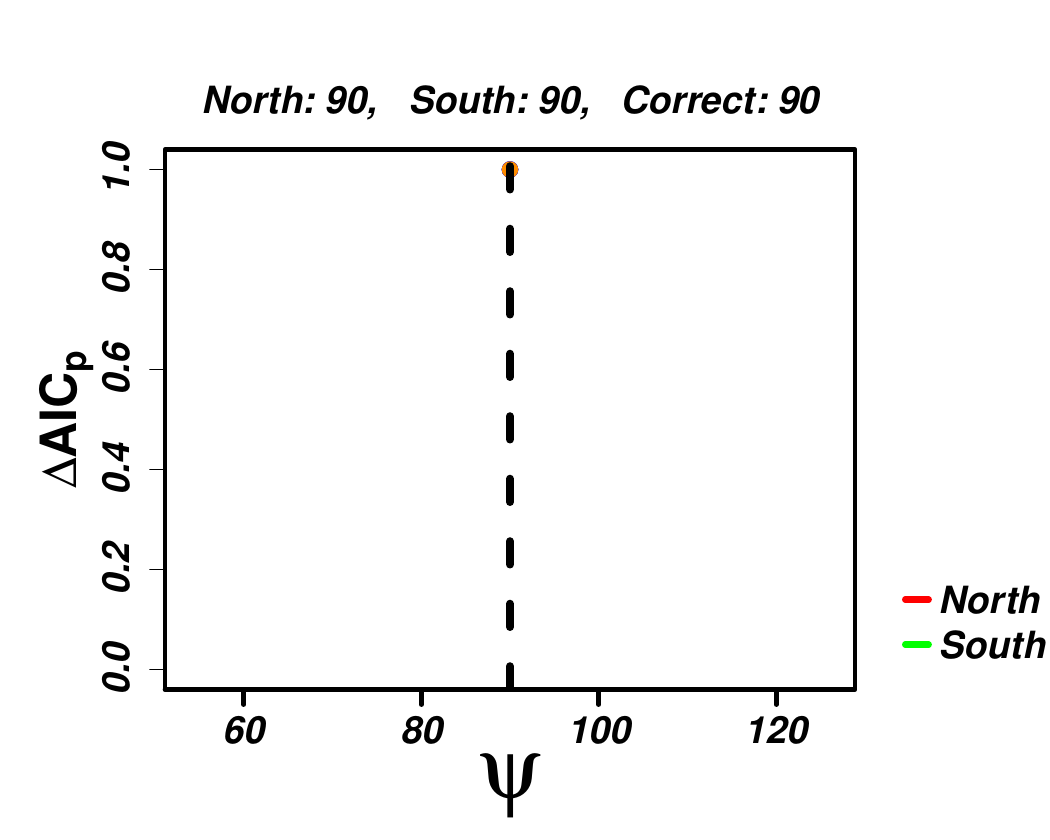}}

    \caption{1D-plots of AIC$_\text{p}$ (eq.\tild\ref{eq.AIC}) against the viewing angles (from left to right: $\left(\theta,\phi,\psi\right)$) for the four $N$-body projections (From top to bottom row: MIDDLE, RAND, INTERM, MINOR) considered in this work. In each plot, the blue points are individual model runs (though we tested many more models outside the plotted range) while the dashed black line labels the true viewing angles. The red and green lines follow the lower envelope of the best-fit models for each $\left(\theta,\phi,\psi\right)$ value and refer to the two halves of the galaxy ("North" and "South"), which we model separately. For the MINOR projection, we show the models along the principal axes when testing different deprojections as orange points.}
    \label{Fig.incl_recovery}
\end{figure*}

\subsection{Dynamical modeling}  \label{Ssec.dyn}

Having significantly reduced the number of possible viewing angles using photometry only, we now turn to the dynamical modeling in order to further constrain the viewing angles and therefore the galaxy shape. Technical details about how we derive the kinematics derived from the simulation, such as the resolution and the number of fitted bins for each projection, the noise level etc. are discussed in Paper I. \\
Our strategy consists of taking each density distribution, which has survived the deprojection cut-off, and its corresponding set of viewing angles and use it in eq.\tild\ref{eq.SMART} along with different \mbh, \ml\, values and the DM halo density. We sample 10 \mbh\,values linearly spaced in the range $\left[1 \times 10^{10}, 3 \times 10^{10}\right] M_\odot$, while for \ml\, we use $\left[0.6,1.4\right]$ as interval, again sampled with 10 values. 
For the DM halo we assume a profile similar to a \citet{Zhao1996} profile (with $\alpha = 1$) in the scaled ellipsoidal radius $r/r_s$. The free parameters are the inner slope $\gamma$, the scale radius $r_s$ and the density normalisation $C$. Since the $N$-body simulation started with progenitor halos approximated by a Hernquist sphere, we adjust $\beta$ such that the outer logarithmic slope of our halo model equals $(\beta-\gamma) = 4.5$ rather than the canonical NFW value of $(\beta-\gamma) = 3$. We measured the asymptotic slope of the simulated halos directly from the DM particles of the simulation at large radii. Because the DM halo derived from the simulation is triaxial we have the two intrinsic shape parameters p$_\text{DM}$, q$_\text{DM}$ of the DM halo as additional free parameters. Thus, we have a total of 10 parameters to be fitted. For this, we use the NOMAD
software \citep{Audet06, LeDigabel11, Audet17} which automatically explores the parameter space, looking for the model minimizing AIC$_\text{p}$. The best-fit model parameters are the ones where AIC$_\text{p}$ is minimized. \\
In principle, the AIC$_\text{p}$ distribution -- just like $\chi^2$ -- could be used to estimate the uncertainty of the models (masses, viewing angles etc.). We will illustrate this below for the anisotropy. However, in general, we prefer to estimate model uncertainties in a different way for two reasons. Firstly, when dealing with real observations the noise level in the data is often not exactly known but has to be estimated -- which makes AIC$_\text{p}$ or $\chi^2$ distributions uncertain by themselves. Secondly, reliable errors from AIC$_\text{p}$ or $\chi^2$ require to sample the respective distributions in all parameters \textit{completely}. In order to gain efficiency (we only seek to find the \textit{optimal} model) and to get independent from the noise estimate for the data, we split the available kinematical data into two sets, which we label "North" and "South". The data set to which each bin is assigned to gets determined by the bin position with respect to the galaxy's apparent minor axis. In this way, each of the two data sets has roughly the same amount of bins. We model these two data sets independently and estimate the uncertainty of all relevant model parameters from the variance between these two fitting runs. 
Below we will also show the 1D functions 
AIC$_\text{p} \left(\theta\right)$, AIC$_\text{p} \left(\phi\right)$, AIC$_\text{p} \left(\psi\right)$ obtained by minimising AIC$_\text{p}$ over all other parameters to illustrate how well individual parameters can be estimated. For the simulations this is justified since we know the noise level in the ``data'' exactly. We summarize in Tab.\tild\ref{Tab.numbers} the number of tested viewing angles,
those that survived the RMS cut-off and those for which AIC$_\text{p}$ coming from the dynamical model is within 50 + AIC$_{\text{p,min}}$ (''good dynamical models'')\footnote{This AIC threshold allows us to perform a very conservative comparison, since typically a diffence of 10-20 in AIC$_\text{p}$ is already considered significant.}.

\begin{table*}
   
\caption{Summary of the discarding process of possible viewing angles using photometry only. \textit{Col.1:} The particular model (see also Tab\tild\ref{Tab.projections}). \textit{Col.2:} Total number
of deprojections (i.e. different viewing angles) that we tested. \textit{Col.3:} Number of viewing angles left over after the RMS cut-off. \textit{Col.4:} Number of densities that we
dynamically modeled. The difference between \textit{Col.3} and \textit{Col.2} yields the number of viewing angles omitted through the RMS, whereas the difference between
\textit{Col.4} and \textit{Col.3} yields the number of viewing angles omitted because of implausible $p(r)$ and/or $q(r)$ profiles.}

\label{Tab.numbers}

   \begin{tabular}{c c c c}
Model & Number of deprojections & After depro cut-off & Dynamical modeling \\   
\hline \hline
MIDDLE & 1800 & 61 & 10 \\
RAND & 1800 & 75 & 9 \\ 
INTERM & 1800 & 25 & 6 \\ 
MINOR & 1800 & 25 & 6 \\ 
\hline
\end{tabular}
    
\end{table*}

\section{Results and Discussion}

We now turn to the analysis of the results and discuss them in detail, focusing on the viewing angles, anisotropy and shape recovery. The results on \mbh\, and \ml\, are discussed in BN21; for completeness, we report them in Tab.~\ref{Tab.BH_ML}.

\begin{table}
\centering
\begin{tabular}{c c c}
 & |\mbh| & |\ml| \\   
\hline \hline
MIDDLE & $1.67 \pm 0.07$ & $1.04 \pm 0.05$ \\
RAND & $1.67 \pm 0.22$ & $1.09 \pm 0.13$ \\ 
INTERM & $1.78 \pm 0.11$ & $1.05 \pm 0.08$ \\ 
MINOR & $1.56 \pm 0.11$ & $1.05 \pm 0.08$ \\ 
\hline
\end{tabular}

\caption{Recovery precision of \mbh\, and \ml\, for the four projections described in this paper. The \mbh\, values are given in units of $10^{10} M_\odot$. The deviations from the true values $\mbhm = 1.7 \times 10^{10} M_\odot$ and $\mlm = 1.0$ are less than 5\%. Finally, the standard deviations are given by modeling the two sides of each projection.} 
\label{Tab.BH_ML}  

\end{table}

\subsection{Viewing angles recovery} \label{Ssec.inclination}

dN20 showed that photometric constraints alone can shrink the range of possible viewing
  angles significantly because deprojections assuming the wrong viewing angles fit the observations not as good as deprojections near the true viewing angles. Hence, first testing deprojections in one octant and then
  using a cut-off in the RMS achieved between the photometric data and the deprojection model allows to select the best intrinsic densities. In
Fig.\tild\ref{Fig.Maps_depro_SMART} we show how well the photometric data of the merger simulation can be deprojected as a function of the assumed orientation of the line-of-sight. The panels with the correct value of $\psi$ are closely analogous to Fig. 20
of dN20. They show that the deprojection alone helps in reducing dramatically the
range of viewing angles that need to be tested when dynamically modeling
the galaxy.

The naive expectation would be that the additional constraints from the observed kinematics will improve the
viewing angle recovery. The results of the dynamical modelling are 
summarized in
Fig.\tild\ref{Fig.incl_recovery}, where we show AIC$_\text{p}$
(eq.\tild\ref{eq.AIC}) as a function of the three viewing angles
(Fig.\tild\ref{Fig.FOR}) for the four $N$-body projections.
Away from the principal axes (i.e. MIDDLE and RAND)
the viewing angles are well constrained (within $15^\circ$), while for the
two cases where the LOS coincides with one of the principal axes (INTERM and MINOR) 
this is only true for $\phi$ and $\psi$ (which are recovered
correctly). The third angle $\theta$ shows a slightly larger offset ($30^\circ$ for INTERM, $20^\circ$
for MINOR). For these two principal-axis projections the angle
$\psi$ was already fixed from the photometric constraints alone.

To compare the dynamical and photometric results we have also included the best-fit dynanmical viewing angles in Fig.\tild\ref{Fig.Maps_depro_SMART}. In some cases the dynamically determined best-fit $\psi$ differs from the correct value. We included additional panels for these best-fit $\psi$ if necessary. In general, for the \textit{correct} value of $\psi$, our Schwarzschild code can
constrain the LOS position with less scatter than the deprojection
alone -- as expected -- although with no significant improvement on the best-fit
viewing angles.

\subsection{The primary importance of the deprojected shape} \label{Ssec.shape}

%

\begin{figure}
    \centering
    \includegraphics[scale=.2]{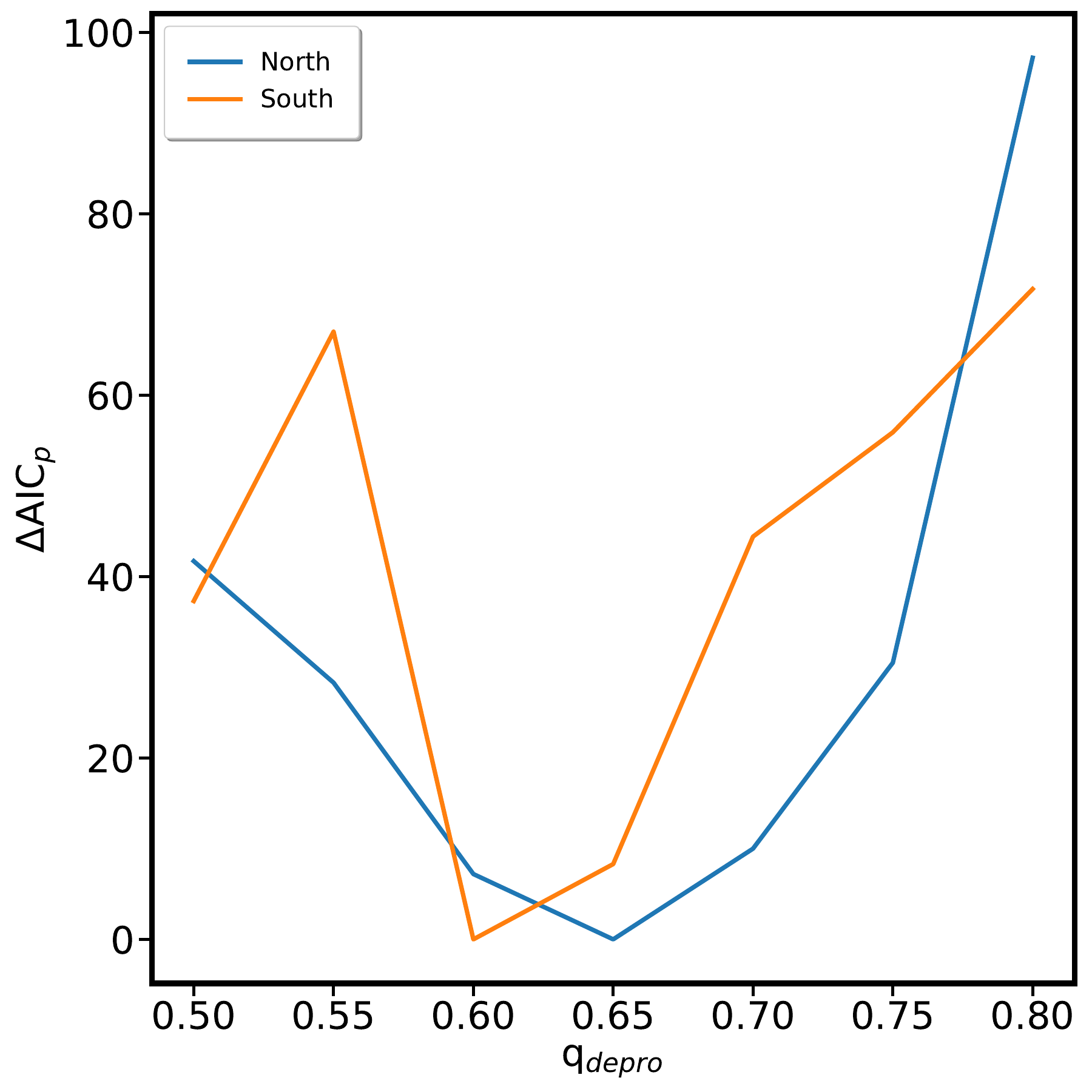}
    \caption{Best-fit AIC$_\text{p}$ values plotted against the corresponding $q$ value used to generate $\rho$. The results refer to the MINOR projection.}
    \label{Fig.AIC_q}
\end{figure}

As already noted, the deprojection of a triaxial body is generically degenerate.
In particular, there exist infinitely many deprojections even at a given viewing angle.
For example, the flattening $q(r)$ is completely unconstrained when the LOS coincides with the intrinsic minor axis. The reason is that the intrinsic density distribution along the z-axis (the LOS in this case) is photometrically simply not accessible in this case (and similar for the other principal axes). However, our deprojection code SHAPE3D allows to probe deprojections with different intrinsic shapes at the same viewing angle. So, for the MINOR projection we constructed a set of deprojections all assuming the minor-axis as LOS but with different intrinsic $q(r)$. In Fig.~\ref{Fig.AIC_q} we show for each constant $q(r)$ value (see Sec.~\ref{Ssec.photometry}) in the
range $\left[0.5, 0.8\right]$, the corresponding best-fit
$\Delta$AIC$_\text{p}$ values for the two modeled galaxy halves. The
minima are located at $q = 0.65$ and $q = 0.6$, in agreement with the
true profile in the region where the galaxy becomes triaxial (see
Fig.~\ref{Fig.pq_true}). In other words, the kinematic constraints are sufficient to identify the correct intrinsic shape out of the set of deprojections with different $q(r)$. In fact, the constraints on the shape are quite strong: the $\Delta$AIC$_\text{p}$ changes by more than $\Delta$AIC$_\text{p} > 40$ over the sampled shape interval. For comparison, in the case of the MINOR projection, a change of $\Delta$AIC$_\text{p} \sim 40$ corresponds to a change of $\theta \sim 40^\circ$. The constraints on the shape are therefore quite significant, in particular for the viewing angle recovery. 

In Fig.\tild\ref{Fig.incl_recovery} the improvement in AIC$_\text{p}$ due to the inclusion of the above described deprojections with different $q(r)$ is indicated by the orange points. As expected,  
including additional deprojections with different intrinsic flattenings for this
orientation significantly improves the results of the viewing angle recovery. In fact, they allow to
recover the correct galaxy viewing angles in the "southern" part of the
galaxy. 

These results imply that the constraints from modern kinematic data may in many cases be sufficient to discriminate between different deprojections at the same viewing angle. For a successfull viewing angle recovery flexible deprojection tools that allow to probe different intrinsic shapes at a given viewing angle are therefore vital. In particular, our results suggest, that if we would try to also optimise the deprojected shapes at other LOS orientations, the viewing angle recovery with the dynamical models would improve. However, the precision that we can achieve with our fiducial deprojections in terms of the mass recovery (see Paper I) and in terms of the recovery of the intrinsic shapes and orbital anisotropies (see below) suggest that this is not necessary and that the viewing angles themselves are only of secondary importance for the dynamical modelling.

\begin{figure*}

\subfloat[MIDDLE]{\includegraphics[height=.35\paperheight]{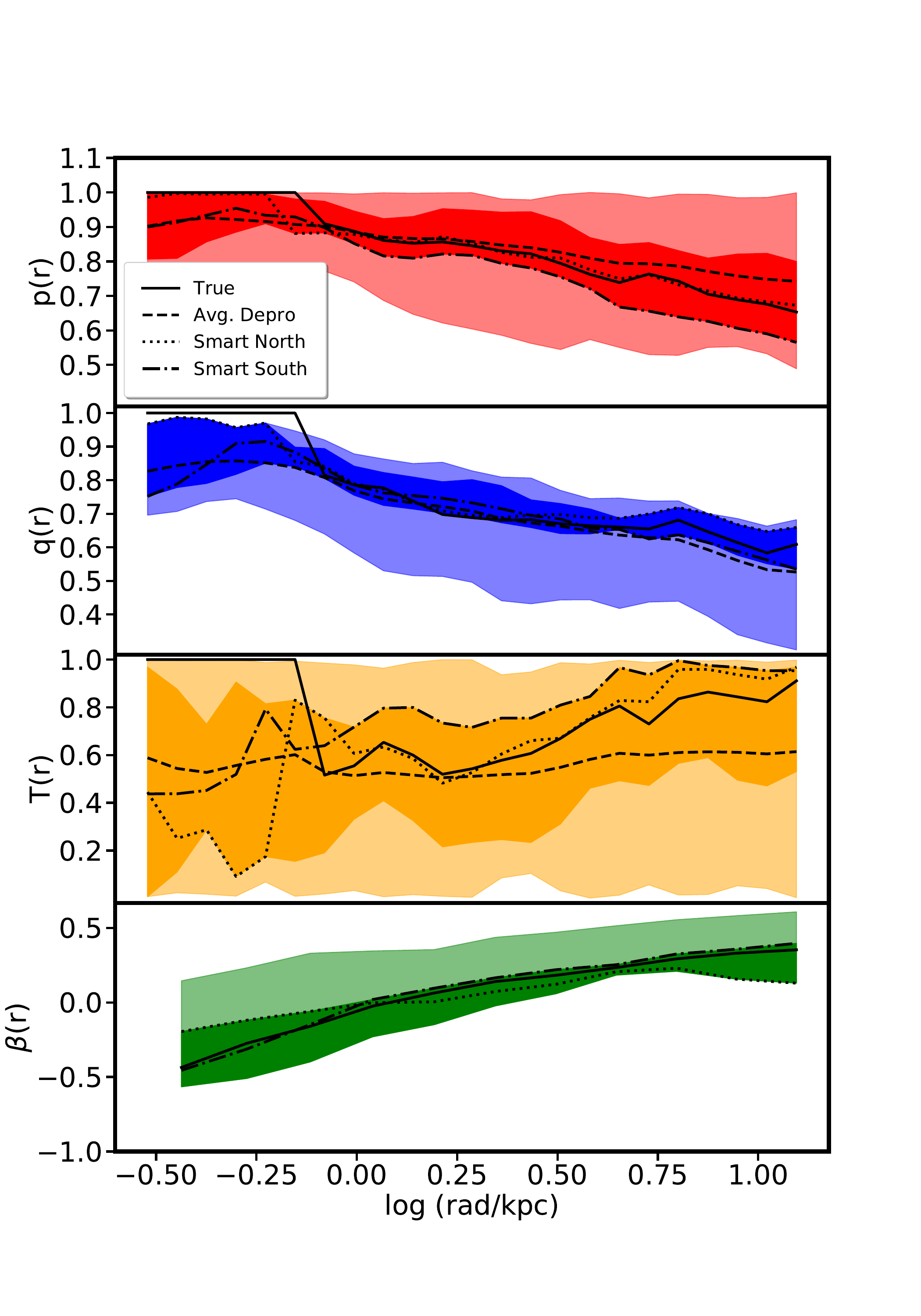}}
\subfloat[RAND]{\includegraphics[height=.35\paperheight]{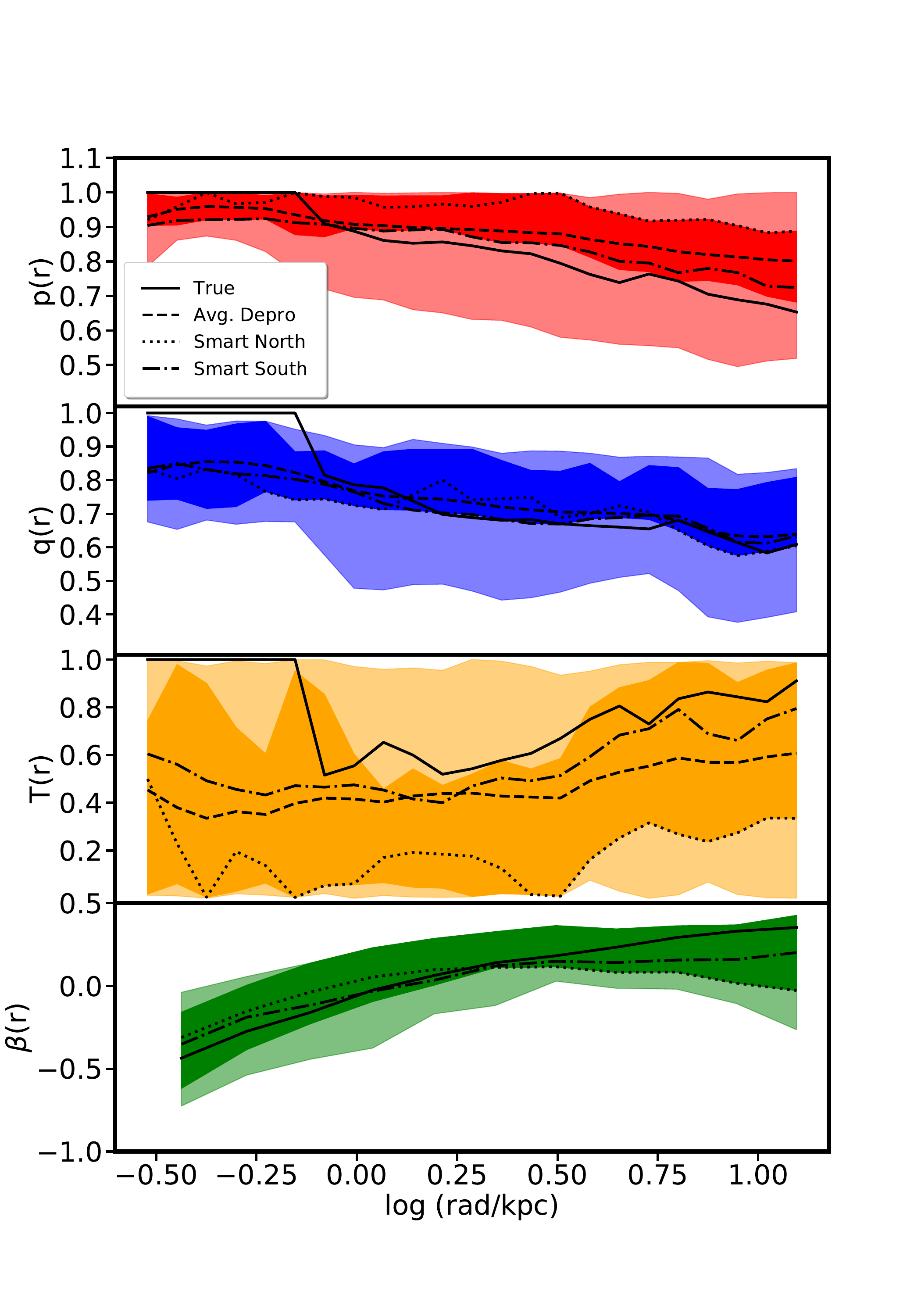}}

\subfloat[INTERM]{\includegraphics[height=.35\paperheight]{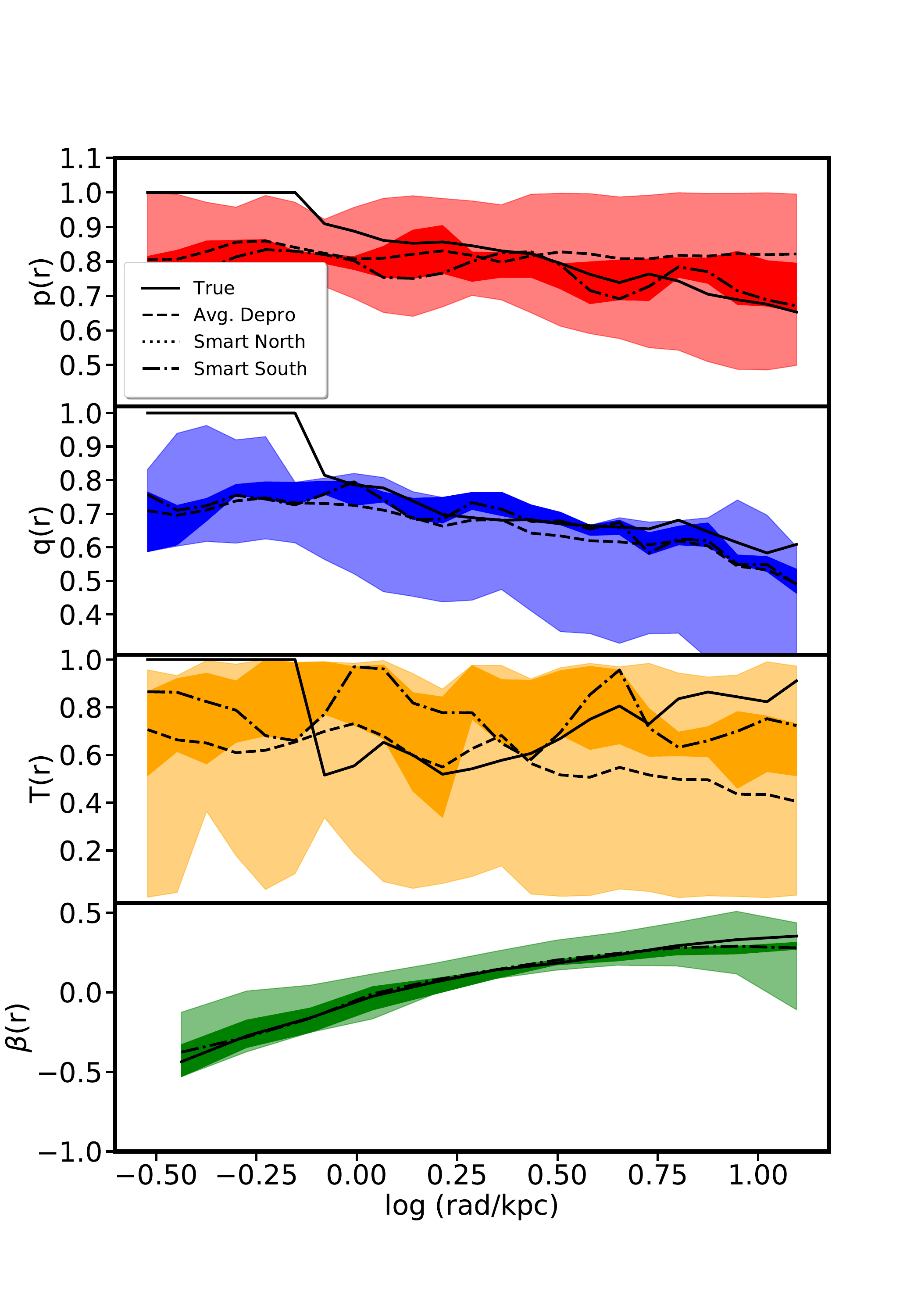}}
\subfloat[MINOR\label{Fig.pq_Beta_minor}]{\includegraphics[height=.35\paperheight]{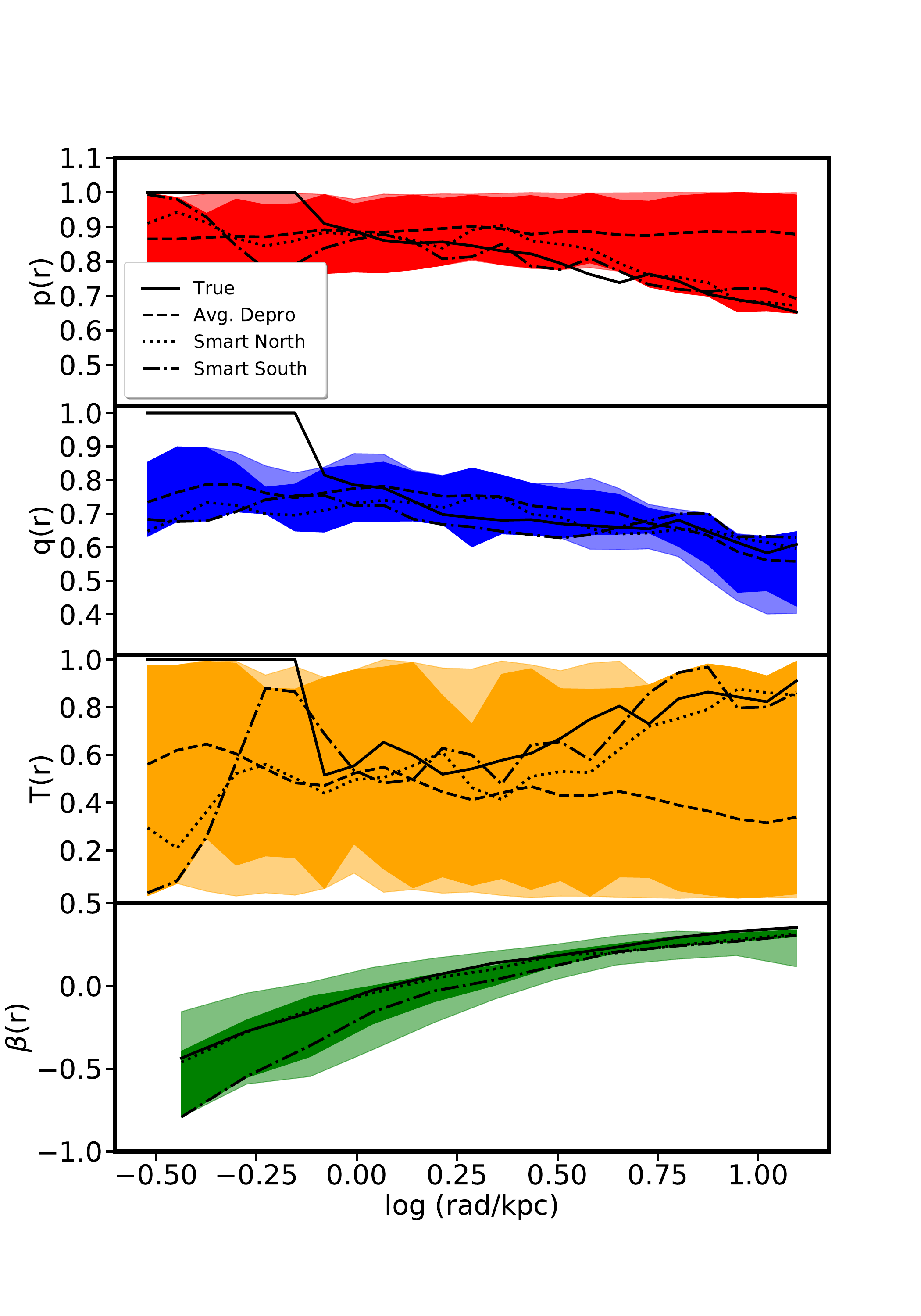}}

    \caption{For each of the four projections we show, from top to bottom panels, $p(r), q(r), T(r)$ and $\beta(r)$ intervals that we get considering every acceptable deprojection (lighter color) or only those deprojections for which the dynamical model yields an AIC$_\text{p}$ $\leq \text{AIC}_{\text{p,min}} + 50$ (darker color). The solid lines show the correct profiles, whereas the dashed lines show the average profile that we get among all good deprojections. Finally, the dotted and dash-dotted lines are the best-fit profiles from the dynamical models.}
    \label{Fig.pq_Beta}
\end{figure*}

\subsection{Shape recovery} \label{Ssec.shaperecover}

The orientation parameters and the intrinsic shape are
intimately connected. However, there is no generic simple connection for \textit{triaxial} objects in general -- due to the high degree of degeneracy in the deprojection (cf. last Sec.~\ref{Ssec.shape}). In many currently used deprojection methods a close connection between shape and orientation is nevertheless imprinted through additional assumptions made upon the intrinsic structure of galaxies. Now we test how well we can recover the intrinsic shape of the merger simulation from the four projected mock data sets.


Our
  main results from the dynamical modelling are:

\begin{itemize}
    \item Our results are accurate: we get $\Delta p$ and $\Delta q$ $\lesssim$ 0.1 for each of the four projections;
    \item The kinematical data help in reducing the scatter of the recovered profiles;
    \item The results are best when the LOS is not one of the principal axes.
\end{itemize}

This is summarized in Fig.\tild\ref{Fig.pq_Beta}. For the purely photometric results, we compare the shape {\it profiles} yielded by the
  deprojections using a similar approach as in
  \citet{dN22}, i.e. we consider every deprojection that is qualified as good according to the criteria given in \citet{dN22} and calculate
  the average $<p(r)>$, $<q(r)>$ among them. The range encompassed by these 'good' deprojections is shown by
  light-coloured regions in Fig.\tild\ref{Fig.pq_Beta}, the average profiles are
  shown as dashed lines. We calculate
  the differences between $< p(r) >$, $< q(r) >$ and the true profiles
  at all radii and then average these, reporting the results in
  Tab.~\ref{Tab.deltapq}.  We observe that the shape is best recovered
  for MIDDLE, with $\Delta p, \, \Delta q < 0.06$.  Similarly good is the recovery of the rounder RAND
  projection. This is somewhat
  expected, since when the viewing angles are located between the
  principal axes, SHAPE3D performs optimally in recovering the
  intrinsic shape if the viewing angles are known
  \citep{dN20}. 
 
The results for the principal-axis projections INTERM and MINOR are slightly worse, which is again expected
given the fact that the intrinsic shapes are less
constrained\footnote{For example, along the intermediate axis, the
  fact that we choose $P = 0.8$ for the $\rho$-grid implies that
  deprojections close to the intermediate axis will be biased towards
  $p(r) \sim 0.8$ at each $r$.}. Nevertheless, the intrinsic shapes
can be reconstructed photometrically with an accuracy of $\sim0.1$.

For all four projections, the best-fit triaxiality {\it profiles}
deviate from the truth in the central regions (where the simulation is
spherical), but approach the value of one in the outskirts, where the
simulation becomes prolate.

The dynamical modelling improves the shape recovery significantly. This is shown
in Fig.\tild\ref{Fig.pq_Beta} by the dark-coloured
  regions which encompass all dynamical models within $\text{AIC}_{\text{p}} \leq
  \text{AIC}_{\text{p,min}} + 50$. The figure also shows the intrinsic shape profiles of the actual best-fit dynamical model for each projection and each modelled half as dotted and dash-dotted lines,
  respectively.

The darker regions from the dynamical models are in all cases within the larger uncertainty regions derived from the photometric constraints alone (light colors). 
To quantify this for the shape parameters, we average $p(r)$ and $q(r)$ over all models within $\text{AIC}_{\text{p}} \leq
  \text{AIC}_{\text{p,min}} + 50$. The differences between these \textit{average} profiles and the true ones and also the difference between the profiles of the single best-fit models and the true profiles are quoted in Tab.~\ref{Tab.deltapq}. As one can see, adding kinematical
  data to the analysis does improve the precision of the estimates, but not their accuracy.
  Even by choosing a very high threshold $\text{AIC}_{\text{p,min}} +
  50$ to select the favoured dynamical models, the interval embedding
  our 'good deprojections' is narrowed down in most cases,
  delivering smaller scatter. The true profiles are found
  within these intervals or very close to them, with the only
  exception being $p$ for INTERM and $q$ for MINOR in the central
  regions. This is expected, since $p$ (for the INTERM
  projection) and $q$ (for the MINOR projection) are hidden by the
  LOS (cf. Sec.~\ref{Ssec.shape}).
  
 \subsection{Anisotropy recovery} \label{Ssec.anisotropy}

The 3D intrinsic shape measured in terms of the $p(r)$ and $q(r)$
profiles and the orbital structure -- the anisotropy profile
$\beta(r)$ of a galaxy are related to each
other through the Tensor Virial Theorem, although the shape does not uniquely determine the orbital anisotropy. Therefore, even the good shape recovery that we discussed in the last Sec.~\ref{Ssec.shaperecover} does not guarantee that the orbital structure is well recovered. We quantify the anisotropy from the internal velocity moments yielded by our
triaxial Schwarzschild code. They are computed as the quadratic mean of $\sigma_r$, $\sigma_\theta$, $\sigma_\phi$ in spherical shells. 

The main results from the dynamical modelling are very similar as for the shape recovery: our results are accurate and we get $\Delta \beta$ $\lesssim$ 0.1 for each of the four projections (Fig.~\ref{Fig.pq_Beta} and Tab.~\ref{Tab.deltabeta}). In particular, we recover the expected tangential bias in the
  central regions for all four projections very well together with the
  general trend towards radial anisotropy at larger radii. The largest
  deviations occur at the smallest and the largest radii,
  respectively, which is expected\footnote{At large radii, because
    near the edge of the field of view (FoV) the constraints on the
    orbital structure become weaker since ever more orbits have
    apocentres outside the region constrained by kinematical data. The
    same is true at small radii, where the finite resolution (mostly
    of the simulation) weakens the constraints on the orbit
    model.}

\subsection{Model advancements}
\begin{figure}
    \centering
    \includegraphics[scale=.2]{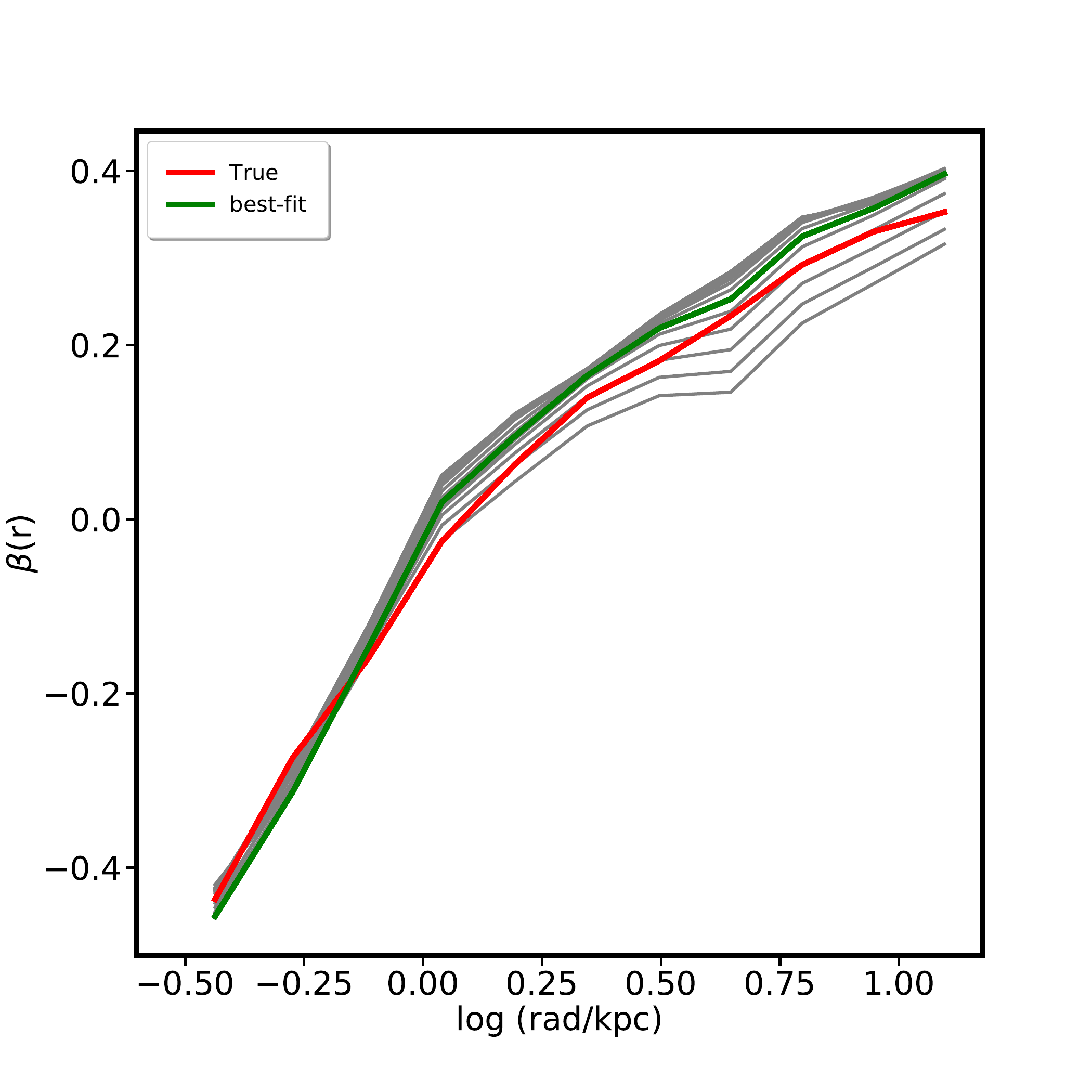}
    \caption{For the best-fit mass model and orientation of the southern half of the MIDDLE projection we show all $\beta$(r) profiles for every smoothing strength $\alpha$ that leads to an acceptable kinematic fit ($\chi^2/N_\mathrm{data}<1$). The red line is the true profile from the simulation, while the green line is our fiducial model derived from the smoothing optimisation using AIC$_p$. All other profiles are plotted as grey lines. We see that the smoothing optimisation is an important factor for the improved accuracy of dynamical models that we report here.}
    \label{Fig.JensThomas}
\end{figure}

The analysis performed in this paper shows that the combination of our triaxial deprojection and dynamical modeling codes enables us to recover the intrinsic shape and anisotropy of the galaxy with good accuracy, meaning that intrinsic degeneracies in both the deprojection problem and in the determination of the orbital dynamics do not play a significant role (see Sec.~\ref{Ssec.summary} below). This is true at least for the setup that we have chosen (integral-field-type data coverage and usage of the entire LOSVDs as constraints), for the given simulation (with a realistic formation scenario and intrinsic shape/orbital structure) and for our newly developed codes. An important contribution comes from SHAPE3D, which allows to narrow down the range of possible viewing angles significantly. Within the variation of the intrinsic shapes among the remaining viewing angles, this already allows to determine the shape of the simulation with good accuracy. A posteriori, this verifies that SHAPE3D alone can be used to make inferences about the intrinsic shapes of real galaxies based on photometric data alone -- at least in a statistical sense \citep{dN22}. \\
BN21 showed that with noiseless data and without the uncertainty introduced by the deprojection step the anisotropy of the same $N$-body simulation can be reconstructed within $\Delta \beta = 0.05$ with our triaxial orbit code. The larger errors here come from the realistic amount of noise in the kinematic data and from the uncertainties intrinsic to the deprojection. However, the anisotropy recovery is still very good. The fact that we have chosen to go up to $\text{AIC}_{\text{p,min}} + 50$, which is very conservative, and that we are able to recover the correct $p(r)$, $q(r)$ with an accuracy typically smaller than 0.085 and reproduce the correct trend of $\beta(r)$ regardless of where the correct viewing angles are, provides a further indication of the stability of our dynamical models and the negligible role of degeneracies when using a setup as described here (see also Sec.~\ref{Ssec.summary}). \\
\citet{Mathias21} have shown that the optimisation of the mass and orientation parameters of Schwarzschild fits is a model selection problem rather than a simple parameter optimisation. Intimately connected to this is the fact that the effective number of parameters m$_\text{eff}$ that a Schwarzschild Fit consists of is variable from model to model. The different number of DOFs, which depends on the particular model (in our case the chosen potential), can bias the results if one uses a $\chi^2$ optimization method. We discuss the importance of the correct model selection for the mass models in Paper I. In a similar way, the optimisation of any smoothing penalty can also be performed using a model selection \citep{Jens22}. To demonstrate the importance of the smoothing optimisation, we show in Fig.~\ref{Fig.JensThomas} for the MIDDLE projection the true $\beta$(r) profile, along with various profiles that we get for the southern half of the merger by just varying the smoothing strength $\alpha$ yet keeping the mass distribution and orientation fixed. We see that even in the same mass distribution different $\alpha$ values may yield anisotropy profiles with deviations of up to 30\% from the true one. And this even though we only show $\beta$ profiles that lead to formally acceptable fits to the data (i.e. $\chi^2/N_\mathrm{data}<1$). An optimal choice of the smoothing is hence necessary to reach the accuracy that we report here.

\subsection{Summary and Discussion} \label{Ssec.summary}
To summarize, the combination of our deprojection and dynamical
modeling recovers the correct shape and anisotropy of the simulated galaxy with deviations
$\lesssim 0.1$. As we show in Paper I the mass recovery has a similar accuracy of about 10 percent. This is not surprising as the masses can only be recovered with high accuracy when the orbital structure is correct and vice versa. The viewing angles turn out to be the most uncertain properties with an accuracy of about $\sim30^\circ$.

BN21 have shown that the anisotropy can be recovered very robustly from kinematic data similar to the one used here (full non-parametric LOSVDs and two-dimensional spatial coverage) when the LOS is given. In Sec.~\ref{Ssec.shape} we have seen that the shape can be recovered very robustly when the LOS is given. Moreover, since in these models the anisotropy was not held fixed and because we know that the best-fit models have the correct anisotropy it is quite straightforward to conclude that kinematic data of the kind used here contain enough information to constrain shape and anisotropy (and mass) together at a given LOS. A little more surprising is the fact that the full modelling of the N-body reveals that both, anisotropy and shape, can even be recovered robustly when the orientation of the LOS is a bit uncertain. One interpretation would be that the constraints on the viewing angles, the shape and the anisotropy do not interfere strongly with each other. And an observation in favour of the possibility that at least shape and anisotropy are not strongly entangled is the fact that the predicted uniform central anisotropy structure in massive (triaxial) ellipticals, i.e. the systematic change from central tangential anisotropy to outer radial anisotropy around the core radius, has already been demonstrated with axisymmetric models \citep{Jens14}. To a certain degree our tests suggest that something similar is true for shape and viewing angles: that the shape constraints force the model towards the correct deprojection even if the viewing angles are not correct. However, we have also seen that this critically depends on whether or not a deprojection with an appropriate shape is among the candidates or not. Flexible deprojection tools like SHAPE3D are therefore important. All in all it has emerged that for the recovery of the masses, intrinsic shapes and orbital structure the correct viewing angles are only of secondary importance. This is probably related to the fact that the merger simulation studied here and massive elliptical galaxies as well are not very strongly flattened. 



\subsection{Bias vs. scatter}
A very important point that can now be addressed is whether or not one needs to go through the dynamical modeling in order to obtain acceptable estimates of the galaxy shape profiles. From Tab.~\ref{Tab.deltapq} we see that the \textit{bias} from the true shape of the simulation which we obtain can only be barely improved (if at all) using kinematical information. Therefore, our conclusion is that the photometric information suffices to obtain a robust estimate (within 0.1) of the correct galaxy shape when considering the average over all orientations as best-fit guess. This approach is exactly what \citet{dN22} used to derive shapes of Brightest Cluster Galaxies (BCGs). \\
Nevertheless, the kinematical information helps in reducing the \textit{scatter}, as it can be seen from both Fig.~\ref{Fig.pq_Beta} (the darker regions are narrower) and Tab.~\ref{Tab.deltapq}. Therefore, the conclusion here is that if one simply wants to obtain an estimate of the galaxy shape, then simply deprojecting the surface brightness profile is enough, but in order to make these estimates more robust, it is still preferable to go through the dynamical modeling. \\ 
Finally, our findings suggest a possible approach for the dynamical modeling of the galaxy, which is needed to determine the orbit distribution along with the mass parameters. In fact, one could model \textit{only one} light density, the one with $p, q$ profiles closest to $< p >, < q >$ and only use this density for the dynamical model. 

\subsection{Comparison with previous studies}

Another study focusing on the recovery of the intrinsic shape of a galaxy using both photometric and kinematic information is \citet{VDB09}. In this work, the authors show that the shape of a triaxial Abel model with constant $p(r) = 0.9$ and $q(r) = 0.77$ projected at $\left(\theta, \phi\right) = \left(60, 60\right)^\circ$ (thus at similar viewing angles compared to our MIDDLE and RANDOM projections) can be well constrained only if the galaxy shows significant rotation both in the central and in the outer regions. Our work shows that for the slowly rotating $N$-body simulation, $\Delta p$, $\Delta q$ $\lesssim$ 0.1 regardless of the photometry and the correct viewing angles. Another improvement with respect to \citet{VDB09} is that in their case when a round, slow-rotator is considered, at almost every viewing angle a solution with recovered $p$, $q\,\sim$0.1 away from the true value (thus very good) can be found, while in our case even for the roundest projection (RANDOM) as well as for those without twists (INTERM and MINOR) our estimates are more accurate and we are able to exclude most of the viewing angles. \\
The triaxial code used by \citet{VDB09} has been used in \citet{Jin19} to recover the intrinsic shapes of nine simulated early-type galaxies from the Illustris simulations (three of which are triaxial). These galaxies are more similar to our $N$-body simulation since they have $p(r)$ and $q(r)$ profiles which are not constant as a function of radius. Here, average deviations of 0.07 in $p$ and 0.14 in $q$ (but with deviations as large as 0.25 in $q$) are found. Nevertheless, only four (for $p$) and one (for $q$) of the nine galaxies they consider show deviations smaller than 0.1, and the anisotropy profiles of these galaxies are also recovered with a lower accuracy (read their Figure 12). Moreover, \citet{Quenneville22} have recently shown that the triaxial code used for the analysis \citep{VDB08} did not project orbits correctly, which may lead to a substantial bias in mass and shape parameters (but see \citealt{Thater22}). \\

Clearly, an exact comparison between these different works is not
possible for a variety of reasons (e.g. \citealt{VDB09} and
\citealt{Jin19} use the MGE as deprojection code), but given the
results presented here our methodology appears superior.




\begin{table*}
   
   \begin{tabular}{c l c c}
Projection & Profile & $< \Delta p > \pm \, \sigma_{\Delta p}$ & $< \Delta q > \pm \, \sigma_{\Delta q}$ \\   
\hline \hline
MIDDLE & AVG ORIENT PHOT & 0.046 $\pm$ 0.084 & 0.059 $\pm$ 0.071 \\
       & AVG ORIENT PHOT + KIN & 0.033 $\pm$ 0.057 & 0.045 $\pm$ 0.042 \\
       & BF SMART & 0.038 $\pm$ 0.034 & 0.044 $\pm$ 0.042 \\
       & & & \\
RAND   & AVG ORIENT PHOT & 0.069 $\pm$ 0.082 & 0.062 $\pm$ 0.089\\
       & AVG ORIENT PHOT + KIN & 0.091 $\pm$ 0.048 & 0.098 $\pm$ 0.057 \\
       & BF SMART & 0.088 $\pm$ 0.055 & 0.070 $\pm$ 0.022 \\
       & & & \\
INTERM & AVG ORIENT PHOT & 0.091 $\pm$ 0.102 & 0.106 $\pm$ 0.067 \\
       & AVG ORIENT PHOT + KIN & 0.088 $\pm$ 0.035 & 0.086 $\pm$ 0.025 \\
       & BF SMART & 0.088 $\pm$ 0.000 & 0.095 $\pm$ 0.000\\
       & & & \\
MINOR  & AVG ORIENT PHOT & 0.105 $\pm$ 0.086 & 0.085 $\pm$ 0.055 \\
       & AVG ORIENT PHOT + KIN & 0.086 $\pm$ 0.082 & 0.084 $\pm$ 0.050 \\
       & BF SMART & 0.053 $\pm$ 0.023 & 0.101 $\pm$ 0.023 \\
\hline
\end{tabular}
    
\caption{Estimates of the mean deviations of the recovered $p(r), q(r)$ profiles from the correct ones from the simulations along with their RMS. \textit{Col. 1:} The projection. \textit{Col. 2:} Whether we consider the average among all deprojections (AVG DEPRO PHOT, light region of Fig.~\ref{Fig.pq_Beta}), among all deprojection for which AIC$_\text{p}$ $\leq \text{AIC}_{\text{p,min}} + 50$ (AVG DEPRO PHOT + KIN, dark region of Fig.~\ref{Fig.pq_Beta}) or the best-fit solution from the dynamical modeling averaged over the two galaxy halves (BF SMART). \textit{Cols. 3-4:} Differences between recovered and correct profiles, averaged on all radii for which we have kinematical data. For the INTERM projection the two best-fit solutions yielded by SMART are the same.  }
\label{Tab.deltapq}

\end{table*}

\begin{table*}
   
   \begin{tabular}{c c c c c}
& MIDDLE & RAND & INTERM & MINOR \\   
\hline \hline
$< \Delta \beta > \pm \, \sigma_{\Delta \beta} $ & 0.067 $\pm$ 0.077 & 0.114 $\pm$ 0.048 & 0.023 $\pm$ 0.000 & 0.078 $\pm$ 0.076 \\
\hline
\end{tabular}
    
\caption{Similar as Tab.~\ref{Tab.deltapq} but with the recovered $\beta(r)$ profiles from the true ones considering the two best-fit models for each galaxy half.}
\label{Tab.deltabeta}

\end{table*}

\section{Conclusions}
We have investigated how well the viewing angles, intrinsic shape and orbital structure of triaxial galaxies can be recovered by employing a novel approach to an $N$-body simulation with high resolution. For the first time we combine our newly developed codes for the modelling of triaxial galaxies: (i) our new semi-parametric triaxial deprojection routine SHAPE3D \citep{dN20} which allows to probe degeneracies of deprojections at the same viewing angle and to shrink the region of possible orientations of a galaxy purely based on photometric data; (ii) our new triaxial orbit superposition code SMART \citep{Bianca21} which exploits the entire kinematic information contained in non-parametrically sampled LOSVDs and uses a 5D orbital sampling to represent all orbital shapes in galaxy centers; (iii) our new model selection methods which allow to adaptively optimise the smoothing for each trial mass model/orientation and overcomes potential biases in $\chi^2$-based approaches.

We tested projections along four representative viewing directions of
this galaxy. We exploit the uniqueness of our deprojections for a
given set of viewing angles and show that the region of possible
viewing directions can be significantly reduced relying solely on the
deprojections.  Using the recovered luminosity densities as input for
our triaxial Schwarzschild code, we determine the correct galaxy
viewing angles to within 15$^\circ$ for MIDDLE and RANDOM, while
  for INTERM and MINOR $\theta$ is 30$^\circ$ and 20$^\circ$ off,
  respectively, but $\phi$ and $\psi$ are perfectly recovered. This
translates to robust estimates of the galaxy intrinsic shape
profiles $p(r)$ and $q(r)$. In two cases where the LOS lies far away
from the principal axes (MIDDLE and RAND), SHAPE3D provides the
correct galaxy shape profiles within 0.1. The same is found for $p(r)$
for the MINOR projection and for $q(r)$ for the INTERM projection.
For the MINOR case, where the $q(r)$ profile cannot be recovered since hidden by the LOS,
we tested different deprojections with different intrinsic shapes,
improving the AIC$_\text{p}$ values and recovering the
correct viewing angles. Thus, even if the best-fit angles do not change,
this exercise leads to an improvement in the quality of the fit
anyway, and should be repeated every time one finds a galaxy whose
photometry is compatible with a deprojection along one of the
principal axes. Moreover, in Paper I we have shown that the best-fit
models also yield the correct BH mass and \ml\,parameters. \\ The
anisotropy parameter $\beta$ shows for each single projection the
tangential bias in the central regions, expected to be generated from
SMBH core scouring. It is significant that this is true even along the
principal axes where either $p(r)$ or $q(r)$ cannot be recovered,
showing the robustness of our Schwarzschild code in recovering the
correct velocity moments and the correct orbital structure of the
simulation. On the other side, the fact that even if the angles are
not exactly correct the $p(r)$ and $q(r)$ profiles are well recovered
(and so are the mass parameters, see paper I) hints at the possibility
of needing a very low number of deprojections for the dynamical
models, therefore reducing the parameter space to be sampled. All that
is needed would be to analyze the favoured deprojections, select
representative $p(r)$ and $q(r)$ profiles and pick up only one
deprojection for each pair of profiles. Finally, our results point out
that the known intrinsic photometric and kinematic degeneracies do not
prohibit a precise and accurate reconstruction of the intrinsic
structure of a triaxial galaxy. In our models, key ingredients are the
non-parametric analysis of the photometric and kinematic data and
advancements in the orbit sampling and model selection. All these
novel improvements will be used in forthcoming works when we will
dynamically model real massive galaxies. \\

\section*{Acknowledgements}
Computations were performed on the HPC systems Raven and Cobra at the Max Planck Computing and Data Facility. \\
This research was supported by the Excellence Cluster ORIGINS which is funded by the Deutsche Forschungsgemeinschaft (DFG, German Research Foundation) under Germany's Excellence Strategy - EXC-2094-390783311. We used the computing facilities of the Computational Center for Particle and Astrophysics (C2PAP).

\section*{Data Availability}
The data underlying this article will be shared on reasonable request to the corresponding author.
	
\bibliographystyle{mnras}
\bibliography{bibl}



\bsp	
\label{lastpage}
\end{document}